\documentclass[11pt,a4paper]{article}
\pdfoutput=1

\usepackage{jcappub}
\usepackage{amsmath}
\usepackage{amssymb}
\usepackage{natbib}
\usepackage{graphicx}
\usepackage{lscape}
\usepackage{hyperref}
\usepackage{xcolor}
\usepackage[normalem]{ulem}

\newcommand{\Lya}{Ly$\alpha$}

\newcommand{\figref}[1]{figure~\ref{#1}}
\newcommand{\secref}[1]{section~\ref{#1}}
\renewcommand{\eqref}[1]{equation~(\ref{#1})}

\newcommand{\QN}{\texttt{QuasarNET}}
\newcommand{\RR}{\texttt{redrock}}
\newcommand{\SQ}{\texttt{SQUEzE}}
\newcommand{\kms}{$\mathrm{km~s}^{-1}$}
\newcommand{\psqd}{sq~deg$^{-1}$}

\title{Optimal strategies for identifying quasars in DESI}

\author[a]{James Farr \href{https://orcid.org/0000-0002-9817-533X}{\includegraphics[scale=0.096]{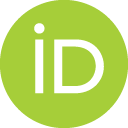}},}

\author[ab]{Andreu Font-Ribera,}
\author[a]{Andrew Pontzen}

\affiliation[a]{University College London, Gower Street, London, WC1E 6BT, UK}
\affiliation[b]{Institut de F\'isica d'Altes Energies (IFAE),
The Barcelona Institute of Science and Technology,
08193 Bellaterra (Barcelona), Spain}

\emailAdd{james.farr.17@ucl.ac.uk, afont@ifae.es}

\abstract{As spectroscopic surveys continue to grow in size, the problem of classifying spectra targeted as quasars (QSOs) will need to move beyond its historical reliance on human experts. Instead, automatic classifiers will increasingly become the dominant classification method, leaving only small fractions of spectra to be visually inspected in ambiguous cases. In order to maximise classification accuracy, making best use of available classifiers will be of great importance, particularly when looking to identify and eliminate distinctive failure modes. In this work, we demonstrate that the machine learning-based classifier \QN{} will be of use for future surveys such as the Dark Energy Spectroscopic Instrument (DESI), comparing its performance to the DESI pipeline classifier \RR{}. During the first of four passes across its footprint DESI will need to select high-$z$ ($z \geq 2.1$) QSOs for reobservation, and so we first assess the classifiers' performance at identifying high-$z$ QSOs from single-exposure spectra. We then quantify the classifiers' abilities to construct QSO catalogues in both low- and high-$z$ bins, using coadded spectra to simulate end-of-survey data. For such tasks, \QN{} is able to out-perform \RR{} in its current form, identifying approximately 99\% of high-$z$ QSOs from single exposures and producing QSO catalogues with sub-percent levels of contamination. By combining \QN{} and \RR{}'s outputs, we can further improve the classification strategies to identify up to 99.5\% of high-z QSOs from single exposures and reduce final QSO catalogue contamination to below 0.5\%. These combined strategies address DESI’s QSO classification needs effectively.}

\arxivnumber{2007.10348}
\begin{document}
\maketitle

\section{Introduction}
\label{sec:Introduction}

The value of spectroscopic surveys hinges on their ability to deliver precise redshift determinations and confident classifications of the objects they observe. Precise redshifts allow clustering analyses to access additional information from modes in three dimensions, while confident classifications ensure that the biases of tracer samples can be accurately assessed. In particular, recent spectroscopic surveys have observed increasingly large populations of quasars (QSOs), using them to study the large-scale structure of the universe in 3D with the direct clustering of QSOs \cite[e.g.][]{Ata:2018MNRAS.473.4773A,Hou:2020arXiv200708998H,Neveux:2020arXiv200708999N} and the clustering of neutral hydrogen via the Lyman-$\alpha$ forest \cite[e.g.][]{deSainteAgathe:2019A&A...629A..85D,Blomqvist:2019A&A...629A..86B,duMasdesBourboux:2020arXiv200708995D}. Traditionally, the joint task of redshift determination and classification of objects targeted as QSOs has relied on visual inspection (VI) by humans, requiring both substantial expertise and time in order to obtain large and reliable sets of classifications. In early surveys, this task was carried out exclusively by VI \cite[e.g.][]{Schmidt:1983ApJ...269..352S}. Subsequently varying degrees of automation were introduced \cite{Hewett:1995AJ....109.1498H,Croom:2004mas..conf...57C}, but VI was still integral to the success of these later surveys, with all spectra being inspected in order to eliminate the $O(5\%)$ classification errors that early automatic pipelines introduced \cite{Croom:2001MNRAS.322L..29C}. This continued with the advent of the Sloan Digital Sky Survey (SDSS) \cite{Gunn:2006AJ....131.2332G,Schneider:2007AJ....134..102S,Schneider:2010AJ....139.2360S}, and in particular the Baryon Oscillation Spectroscopic Survey (BOSS) \cite{Dawson:2013AJ....145...10D} of SDSS-III \cite{Eisenstein:2011AJ....142...72E}. Over the course of its 5 years of operation, BOSS produced three QSO catalogues \cite{Paris:2012A&A...548A..66P,Paris:2014A&A...563A..54P,Paris:2017A&A...597A..79P}, all of which relied on VI to classify spectra accurately. The final BOSS QSO catalogue came from the twelfth SDSS data release (referred to as DR12Q from here on), and consisted of 297,301 visually confirmed QSOs from a ``Superset'' of 546,856 QSO targets.

This enormous VI effort has provided a legacy product of immense value to the community due to the size of its sample and the reliability of its classifications. However, due to the increase in the number of QSO targets observed during the extended-BOSS (eBOSS) programme of SDSS-IV \cite{Dawson:2016AJ....151...44D,Blanton:2017AJ....154...28B}, it was deemed infeasible to repeat such an extensive VI procedure. eBOSS produced two QSO catalogues from the fourteenth and sixteenth SDSS data releases \cite{Paris:2018A&A...613A..51P,Lyke:2020arXiv200709001L}, transitioning towards the use of automatic classifiers \cite[e.g.][]{Bolton:2012AJ....144..144B,Hutchinson:2016AJ....152..205H} in carrying out catalogue construction. The DR14Q and DR16Q catalogues relied on VI for only $\sim$3.7\% and 2.9\% of new spectra respectively, focusing on spectra for which automatic classifiers returned ambiguous results. These automatic classifiers were based primarily on the fitting of spectral templates to each QSO target spectrum, determining a class and redshift from the best fit solutions. This approach treats the problem of classification in a qualitatively different manner to VI, looking to find a minimum $\chi^2$ value over a space of possible templates rather than identifying particular spectral features, as a human would do. In order to provide a complementary solution to template-fitting options, machine learning methods can be employed with the aim of replicating the ``feature detection'' approach of VI. Indeed, the set of VI classifications from BOSS DR12 provides a rich dataset on which to train and test such models. Since this dataset was released, two such tools have been developed, \QN{}~\cite{Busca:2018arXiv180809955B} and \SQ{}~\cite{Perez-Rafols:2020MNRAS.tmp.1953P,Perez-Rafols:2020MNRAS.tmp.1952P}, both of which are able to at least match the performance of template-fitting methods.

Looking to spectroscopic surveys of the future, the number of QSO targets is set to continue its dramatic increase. Over the course of the 5-year main survey of the Dark Energy Spectroscopic Instrument (DESI) \cite{DESICollaboration:2016arXiv161100036D,DESICollaboration:2016arXiv161100037D}, approximately 3.6~million QSO targets will be observed, within which approximately 2.4~million QSOs are expected to be found. This number of objects will require automatic classifiers to further lead the process of classifying QSO target spectra, maintaining extremely high levels of accuracy while minimising any reliance on visual inspection. Ahead of surveys such as DESI, then, it is vital to understand how to make best use of the available QSO classification tools. In particular, with a range of qualitatively different classifiers now available, combining their classifications in order to take advantage of their differing strengths and weaknesses will be of great importance when designing optimal classification strategies.

In this work, we first assess the current landscape of publicly available QSO target spectra, and the tools available to classify them. In \secref{sec:Data and tools}, we provide brief descriptions of the existing data and classifiers alongside a comparison of classifier performance levels. We then look ahead to DESI in \secref{sec:QSO classification in DESI}, setting out the key classification tasks that will need to be addressed during its main survey, and describing broad prioritisations that should be made in each case. In \secref{subsec:Selecting high-z QSOs for reobservation} and \secref{subsec:Constructing a QSO catalogue}, we address the two main QSO classification tasks in turn, considering various strategies to address them and quantifying the effectiveness of each strategy by testing on BOSS DR12Q Superset data.
\section{Data and tools}
\label{sec:Data and tools}
In recent years, the largest sets of QSO spectra have been obtained by the BOSS \cite{Dawson:2013AJ....145...10D} and eBOSS \cite{Dawson:2016AJ....151...44D} programmes, which produced five major QSO catalogues \cite{Paris:2012A&A...548A..66P,Paris:2014A&A...563A..54P,Paris:2017A&A...597A..79P,Paris:2018A&A...613A..51P,Lyke:2020arXiv200709001L}. These catalogues were all constructed in different ways, using different techniques to classify observed spectra of QSO targets. The main BOSS data we use in this work is described in \secref{subsec:BOSS DR12Q Superset data}, while a summary of existing QSO classifiers is given in \secref{subsec:Existing classifiers}.

\subsection{BOSS DR12Q Superset data}
\label{subsec:BOSS DR12Q Superset data}
The final data release from SDSS-III/BOSS was DR12, which included 627,751 spectra from 546,856 objects that were targeted as QSO candidates during the main survey \cite{Ross:2012ApJS..199....3R}. These spectra are each constructed from (on average) four, consecutively-taken, 900~second exposures, which were calibrated and coadded as part of the data reduction pipeline \cite{Bolton:2012AJ....144..144B}. At a later stage of the pipeline, each spectrum was automatically classified and assigned a redshift. However, they were subsequently re-classified via visual inspection (VI) by an expert \cite{Paris:2017A&A...597A..79P} in order to improve the quality of the classification. The full results of VI can be found in the DR12Q ``Superset'' catalogue\footnote{Publicly available at \hyperlink{https://data.sdss.org/sas/dr12/boss/qso/DR12Q/Superset_DR12Q.fits}{https://data.sdss.org/sas/dr12/boss/qso/DR12Q/Superset\_DR12Q.fits}.}, with a detailed explanation of the classification procedure available in ref.~\cite{Paris:2017A&A...597A..79P}. Each object was classified by VI as a star, galaxy or QSO (where data was of sufficient quality), and a VI redshift assigned. These redshifts consisted mostly of values from the automatic pipeline, with corrections applied by the expert during VI as appropriate. A confidence in the results of classification/redshift determination procedure was also given, ranging from 1 to 3 for low to high confidences. This extensive VI effort makes DR12Q Superset spectra ideal for testing the predictions of automatic classifiers, and indeed we make use of this opportunity throughout this work. We consider only spectra from objects that were classified with the highest confidence, and thus can practically consider the VI results on these spectra as ``true'' classifications.

\subsection{Existing classifiers}
\label{subsec:Existing classifiers}

A number of automatic QSO classifiers have been developed in recent years, using a variety of different methods. Here, we outline some of the most prominent examples, dividing them into two sub-groups: those based on the concept of ``template-fitting'' in \secref{subsubsec:Template-fitting classifiers}, and those based on machine learning methods in \secref{subsubsec:Machine learning classifiers}. Finally, in \secref{subsubsec:Performance of classifiers}, we compare the performance of four classifiers on BOSS DR12Q Superset data.

\subsubsection{Template-fitting classifiers}
\label{subsubsec:Template-fitting classifiers}
In recent spectroscopic surveys, the most common spectral classification method has been to find the best fit to a given spectrum from a set of templates, framing the problem as one of $\chi^2$ minimisation \cite{Croom:2001MNRAS.322L..29C,Bolton:2012AJ....144..144B}. This necessitates a sound understanding of the data and its associated errors, as well as a broad set of templates that capture the full variety of features present in the data.

During the BOSS programme, such a classifier was developed as part of the \texttt{idlspec2d} data reduction pipeline \cite{Bolton:2012AJ....144..144B}. This used a set of templates for each class (star, galaxy or QSO) constructed from well-understood spectra measured earlier on in SDSS. To construct the QSO templates, 1,000 QSOs were selected at random from the SDSS DR5 QSO catalogue~\cite{Schneider:2007AJ....134..102S}, enforcing a uniform redshift distribution. Of these, 568 defect-free spectra had been observed by BOSS, and were selected as the QSO template training sample. Due to the BOSS target selection and observation strategies, the redshift distribution of these spectra was strongly weighted towards $z\geq2.2$ (see figure 4 of ref. \cite{Bolton:2012AJ....144..144B}). Template source spectra for all classes were put through a PCA decomposition procedure, with leading principal components retained and used as a linear basis to fit to each DR12 spectrum. For each spectral type, linear combinations of the basis components were fitted to each spectrum at each redshift within a suitable range. From these fits, a best class and a best redshift was determined as the template class-redshift combination that resulted in the lowest reduced-$\chi^2$. As an output, the classifier then provided this best-fit redshift and class, along with a 1$\sigma$ statistical error in the redshift and a \texttt{ZWARNING} flag. This flag was raised for problematic fits, with a variety of possible bits to signify different possible issues. Most notably, the dominant source of warnings was from spectra where the difference in reduced-$\chi^2$ value between the best and the next-best (with velocity difference greater than 1,000~\kms{}) classification-redshift pair was below a threshold of 0.01. These spectra thus had two possible classification-redshift pairs which achieved approximately the same quality of fit, and so the classification and redshift were deemed to be insecure. It achieved little contamination at $z\geq2.2$, but at low $z$ it introduced substantial contamination from stars, as well as missing $\gtrsim5\%$ of QSOs at all redshifts (see \figref{fig:classifier_performance_summary_altcolour}). As such, its accuracy was not considered sufficient for final classifications, motivating the extensive VI programme discussed in \secref{subsec:BOSS DR12Q Superset data}.

Recently, DESI has developed a new template-based classifier named \RR{} to improve upon several aspects of \texttt{idlspec2d}. When comparing the model to the data, \RR{} accounts for the wavelength-dependent spectral resolution of individual fibres. It resamples the model to the wavelength binning of the data rather than using an average resolution and rebinning the data to the model grid, a process which introduces covariances that were not fully modelled in \texttt{idlspec2d}. It also supports the ability to simultaneously fit individual exposures instead of requiring a coadded spectrum, which can also introduce covariances that were not modelled in \texttt{idlspec2d}. \RR{} optionally has the ability to compare the best fits to a suite of galaxy and stellar archetypes, aiming to identify and remove non-physical fits to those classes, such as those constructed to have negative emission lines. It does not, however, include QSO archetypes due to their greater spectral diversity. Finally, \RR{} includes a new suite of galaxy and stellar templates. However, as of version 0.7.2 of \texttt{redrock-templates}\footnote{Publicly available at \hyperlink{https://github.com/desihub/redrock-templates}{https://github.com/desihub/redrock-templates}.} --- as used in this work --- the \RR{} QSO templates are the same as those used by \texttt{idlspec2d}. DESI is developing new QSO templates prior to its main survey, looking in particular to improve the $z<2.2$ QSO performance where the \texttt{idlspec2d} QSO templates' training set are underrepresented, but these new \RR{} templates are not yet available.

\subsubsection{Machine learning classifiers}
\label{subsubsec:Machine learning classifiers}
The extensive VI effort carried out on the BOSS DR12Q Superset data \cite{Paris:2017A&A...597A..79P} provided a large, human-verified set of spectral classifications, ideal for use as training data for machine learning algorithms. Two QSO classifiers have since been developed separately, using very different approaches. Unlike the template-fitting methods described in \secref{subsubsec:Template-fitting classifiers}, these methods classify each spectrum as a binary QSO/non-QSO, and so they cannot fulfil the same all-round capabilities as their template-fitting counterparts. However, for the more specific problem of QSO classification, they are able to offer complementary classification channels to the template-fitting methods described previously, with different areas of strength and weakness.

The first of these classifiers is called \QN{} \cite{Busca:2018arXiv180809955B}, which attempts to mimic human identification of emission lines. \QN{} is a deep convolutional neural network (CNN) classifier, taking a smoothed spectrum as an input before carrying out four layers of convolutions. The output from these convolutions is then passed to a fifth, fully-connected layer, before feeding into a number of ``line finder'' units. Each of these units consists of a fully-connected layer, trained to identify a particular emission line. This is carried out by first dividing each spectrum into a number of wide ``boxes'', equally spaced in log-wavelength. When training, each line finder unit is given a set of binary identifiers for each spectrum which indicate which box its emission line is in (when the line is present). When making a prediction, a line finder unit attempts to replicate these binary identifiers with a set of numbers between 0 and 1. We interpret these estimates as \QN{}'s confidence as to whether a certain line is in a given box, and we take the largest of these as the confidence that the line has been found in the spectrum as a whole. The line finder unit is also trained to predict the offset of the relevant emission line within each box, and we use the offset within the most confident box to obtain a more precise estimate of the line's location. This then allows us to infer a redshift from each emission line. Throughout this work, we train \QN{} models to detect the \Lya{}, C\textsc{iv}~1548, C\textsc{iii}~1909, Mg\textsc{ii}~2796, H$\beta$ and H$\alpha$ lines. These are chosen to ensure that at least two emission lines will be present in the BOSS spectrograph for reasonable QSO redshifts. From \QN{}'s confidences and redshifts for these lines, we carry out classification via a simple procedure: if at least $n_\mathrm{detect}$ lines are found with confidence exceeding a ``confidence threshold'' $c_\mathrm{th}$, then the spectrum is classified as a QSO. The most likely redshift is then taken from the most confidently identified line. Thus, keeping $n_\mathrm{detect}$ fixed, choosing a higher (lower) confidence threshold results in fewer (more) objects being classified as QSOs. Equally, keeping $c_\mathrm{th}$ fixed, increasing (decreasing) $n_\mathrm{detect}$ has the same qualitative effect. The results of varying these two parameters is presented in figure 2 of ref.~\cite{Busca:2018arXiv180809955B}. \QN{} is also trained to identify broad absorption line (BAL) QSO spectra, using the same method as described above for emission lines. It is able to identify BAL spectra with high success rates (see figure 4 of ref.~\cite{Busca:2018arXiv180809955B} and surrounding discussion), though does not currently provide BAL properties such as ``balnicity index'' as more specialised tools do \cite[e.g.][]{Guo:2019ApJ...879...72G}. We do not investigate \QN{}'s BAL performance in this work, instead focusing on its ability to classify QSO spectra via emission line presence and location.

The classifier \SQ{} \cite{Perez-Rafols:2020MNRAS.tmp.1953P,Perez-Rafols:2020MNRAS.tmp.1952P} also attempts to mimic the human process of identifying QSO spectra, by looking for sequences of emission peaks. However, its methods are very different to those of \QN{}. \SQ{} first smooths each spectrum to remove noise features, and searches for emission peaks above a certain significance threshold. Spectra for which no peaks are found are discarded, while those with significant peaks are retained. For each identified peak, \SQ{} then attempts to assign a ``trial identity'' in order to determine a redshift, analogous to the way that a VI expert would attempt to associate an observed peak with a particular emission line. For each possible identity a number of high-level metrics are computed, including the relative strength of the emission line above the continuum, and the slope of the continuum around the line (see equations 1--3 of ref.~\cite{Perez-Rafols:2020MNRAS.tmp.1953P}). These are then passed as features to a random forest classifier. By restricting the features seen by the random forest to high-level metrics, \SQ{} attempts to remove any tendency to learn from spurious features of the training set such as instrumental defects or pipeline reduction errors. The random forest then assesses the validity of each trial peak-identity pair, assigning a ``confidence'' of that pair being a correct identification. Finally, the spectrum is classified as a QSO if the largest of these confidences meets a certain threshold value, which can be chosen according to the specifics of the classification task at hand.

\subsubsection{Performance of classifiers}
\label{subsubsec:Performance of classifiers}
The classifiers described in \secref{subsec:Existing classifiers} all use different techniques, and one would expect variation in performance on different subsets of the data as a result. In order to summarise the performance of the classifiers and provide a comparison, we apply each classifier to BOSS DR12Q Superset data and present the results simultaneously. 

We take the results of the DR12 pipeline from the publicly available data in the \texttt{spAll} file\footnote{\hyperlink{https://data.sdss.org/sas/dr12/boss/spectro/redux/spAll-DR12.fits}{https://data.sdss.org/sas/dr12/boss/spectro/redux/spAll-DR12.fits}.}, restricting to QSO targets by including only data from rows corresponding to object identifiers (column name \texttt{THING\_ID}) listed in the DR12Q Superset file. We obtain \RR{} results by running version 0.14.3 of the code\footnote{\hyperlink{https://github.com/desihub/redrock/releases/tag/0.14.3}{https://github.com/desihub/redrock/releases/tag/0.14.3}.}, and using version 0.7.2 of the templates\footnote{\hyperlink{https://github.com/desihub/redrock-templates/releases/tag/0.7.2}{https://github.com/desihub/redrock-templates/releases/tag/0.7.2}.} on these same QSO target spectra. For both the DR12 pipeline and for \RR{}, we do not classify as QSOs spectra which raised a \texttt{ZWARN} flag\footnote{Such flags are named \texttt{ZWARNING} in \texttt{idlspec2d} but \texttt{ZWARN} in \RR{}. We use the latter name from here on for simplicity.}, as classifications for these spectra are likely to be inaccurate \cite{Paris:2017A&A...597A..79P}. \QN{} results are obtained by applying models trained on 90\% of DR12Q Superset data to all spectra from objects not included within this training sample. We train 10 such models, choosing training/testing splits such that all models have mutually exclusive test sets and thus allowing us to obtain \QN{} classifications for 504,534 DR12Q Superset spectra. \SQ{} results were provided by its developers, and were obtained by applying a model trained on $\sim$3\%\footnote{The convergence of the \SQ{} algorithm's performance with training set size was tested in ref.~\cite{Perez-Rafols:2020MNRAS.tmp.1953P}. Using a training set size larger than this $\sim3\%$ was deemed not to yield any a better-performing model.} of DR12Q Superset data to all objects not included in the training set. Our initial test set consists of all DR12Q Superset spectra for which we have classifications from all four classifiers. To form our final test set, we restrict to those objects for which we have maximally confident VI results to ensure that we are comparing to ``true'' classifications. In total, then, the test set contains 481,201 spectra.

We quantify the classifiers' performance levels via purity and completeness, which we define in the same way as ref.~\cite{Busca:2018arXiv180809955B}:
\begin{equation}
\label{eqn:purity}
    \mathrm{purity} = \frac{\mathrm{number\ of\ correctly\ predicted\ QSOs}}{\mathrm{number\ of\ predicted\ QSOs}},
\end{equation}
\begin{equation}
\label{eqn:completeness}
    \mathrm{completeness} = \frac{\mathrm{number\ of\ correctly\ predicted\ QSOs}}{\mathrm{number\ of\ true\ QSOs}}.
\end{equation}
In order for a classification to qualify as a ``correctly predicted QSO'', we require the classifier to have correctly identified a true QSO's spectrum as that of a QSO, and to have matched the VI redshift with a velocity error
\begin{equation}
\label{eqn:velocity error}
    \Delta v=\frac{c|z-z_\mathrm{VI}|}{1+z_\mathrm{VI}}\leq6000~\mathrm{kms}^{-1}.
\end{equation}
This tolerance level rules out catastrophic failures --- classifications which have mis-identified emission lines, for example --- but does not require a highly accurate determination of the redshift as \QN{} and \SQ{} were not designed to provide such a measurement. When calculating the purity, we also allow galaxy spectra identified by a classifier as QSOs to qualify as ``correctly predicted QSOs'', provided the classifier redshift matches that from VI to within 6,000~\kms{}. This allows for the ambiguity as to whether a spectrum should be classified as a galaxy or QSO when a degree of broad-line emission is observed as a result of AGN activity.

\begin{figure}
\centering
\includegraphics[width=15cm]{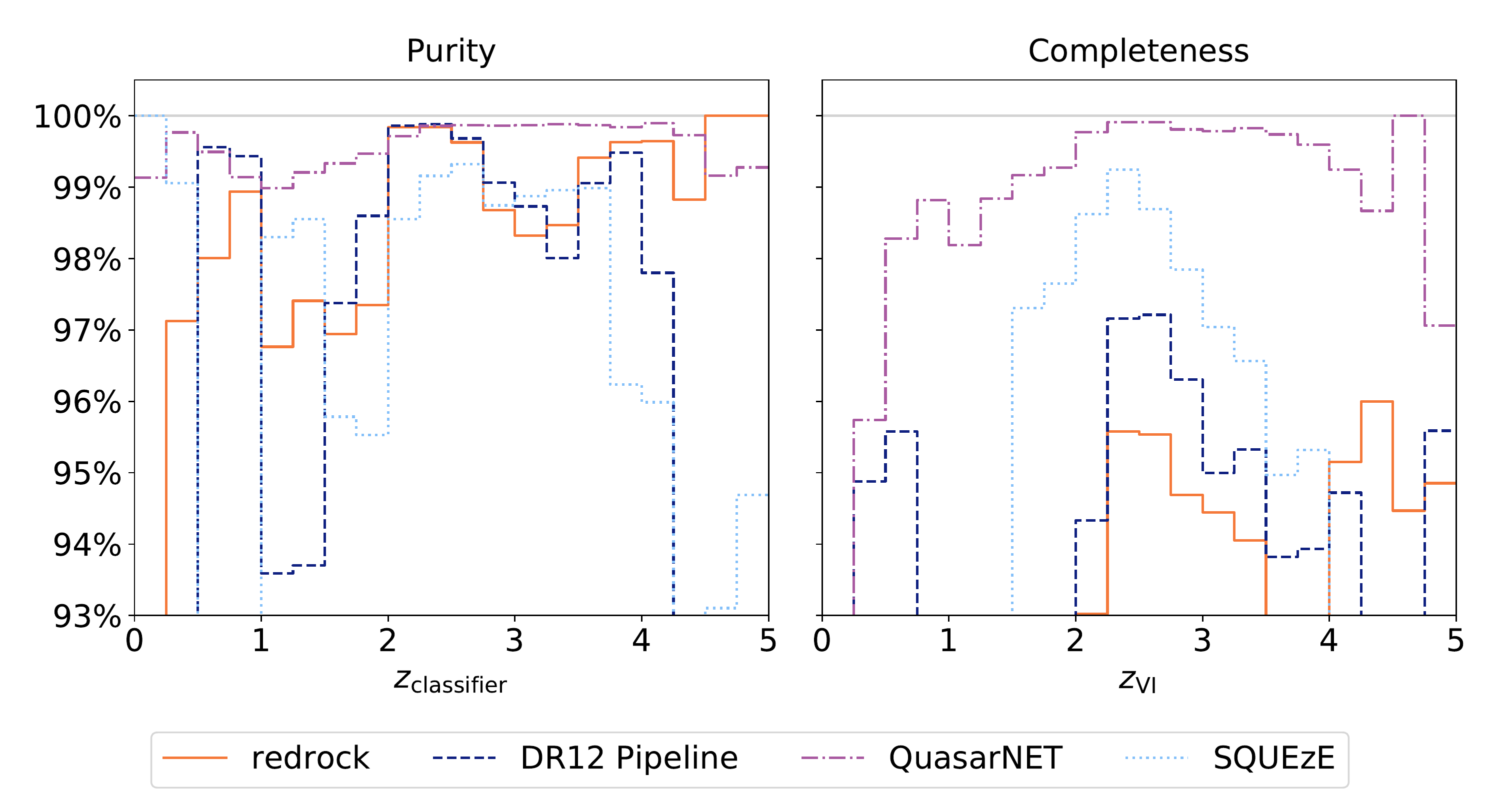}
\caption{Purity as a function of classifier redshift, and completeness as a function of VI redshift for classifications from a selection of current QSO classifiers when applied to a subset of BOSS DR12Q Superset spectra. See equations \eqref{eqn:purity} and \eqref{eqn:completeness} along with surrounding text for definitions of purity and completeness.}
\label{fig:classifier_performance_summary_altcolour}
\end{figure}

The results of our classifications are presented in \figref{fig:classifier_performance_summary_altcolour}, in which the left panel shows purity as a function of classifier redshift, and the right panel shows completeness as a function of VI redshift. It is clear that all classifiers are able to achieve very high levels of purity, with levels exceeding 97\% for the majority of the redshift range. There is a slight dip in purity for all classifiers in the range $1<z<2$, perhaps because neither the \Lya{} nor the narrow O\textsc{iii} 5007 line is present to aid classification in this region (as noted in ref.~\cite{Bolton:2012AJ....144..144B}). This dip is less pronounced for \QN{}, and indeed it appears that \QN{} is able to achieve highest purity across the majority of the redshift range, showing very little variation and achieving $\gtrsim 99\%$ in all bins above $z=0.5$. The other classifiers are able to achieve purities $>99\%$ as well, but are only able to do so in more limited redshift ranges.

Completeness --- shown in the right panel of \figref{fig:classifier_performance_summary_altcolour} --- is also high for all classifiers. It is, however, notably lower than purity and there is more variation in performance, both between classifiers and between redshift bins. \QN{} achieves the highest completeness of the classifiers over the majority of the redshift range, exceeding $99.5\%$ in the range $2<z<4$. This value drops slightly at lower redshifts but still exceeds the other classifiers. A rise in completeness at higher redshifts is also evident for the results from all other classifiers. \SQ{} shows a peak in completeness in the range $1.5<z<3.5$, perhaps related to the presence of the distinctive \Lya{} feature for these redshifts.A sharp increase in completeness is seen at $z\simeq2.2$ for \RR{} and the DR12 Pipeline. This coincides with the redshift value above which the \texttt{idlspec2d} template training spectra increase significantly in number, enabling the template-fitting classifiers to identify a greater range of QSO spectra.

In general, \QN{} is able to out-perform the other classifiers across most redshifts, both in terms of purity and completeness. This reflects the highly flexible nature of its CNN architecture for this ``image recognition'' problem, allowing \QN{} to extract information efficiently and accurately from each spectrum. The other classifiers's methods are not quite as flexible in this respect, with \SQ{} relying on well-defined metrics, and the template-fitting approaches requiring a comprehensive set of templates in order to achieve high performance. Of course, these metrics and templates do allow for relatively easy explanation of individual classifications, whereas \QN{}'s convolutional filters are more difficult to interpret. Nonetheless, the metrics and templates of \SQ{} and \RR{} respectively will be key areas for development when looking to improve the classifiers' future performance.


Whilst \figref{fig:classifier_performance_summary_altcolour} provides a suitable overview of the performance of the different classifiers, it is not helpful in assessing their complementarity. For example, while all classifiers miss $\gtrsim1\%$ of QSOs below $z=2$, the figure does not indicate whether the different classifiers miss the same particular set of QSOs. If the intersection between two classifiers' sets of missed QSOs is small, then combining the classifiers' results in certain ways may enable us to find an even greater proportion of the QSOs in this redshift range. Combining classifications from different classifiers is discussed extensively in \secref{subsubsec:Performance of combined strategies (reobs)} and \secref{subsubsec:Performance of combined strategies (cat)}, taking into account different classification priorities in each case.
\section{QSO classification in the Dark Energy Spectroscopic Instrument}
\label{sec:QSO classification in DESI}

The Dark Energy Spectroscopic Instrument (DESI) will begin its main survey during 2021, and will run for 5 years. DESI will offer significant instrumental upgrades over (e)BOSS; its spectrograph has 5,000 fibres per exposure rather than 1,000, and it will be mounted to a 4m rather than a 2.5m telescope \cite{Gunn:2006AJ....131.2332G,DESICollaboration:2016arXiv161100037D}. This will allow for a substantial increase in the number of objects observed: the forecasted number of QSOs in DESI is $2.4\times10^6$ \cite{DESICollaboration:2016arXiv161100036D}, compared to $7.5\times10^5$ in the final combined (e)BOSS sample \cite{Lyke:2020arXiv200709001L}. There will also be significant differences in the observational methods of DESI compared to (e)BOSS, with objects targeted using different photometric data and with different strategies employed to select from these targets. Both the increased dataset size and the different observing strategies will affect the QSO classification challenges facing DESI, and so new classification methods will be required. 

Ahead of DESI's main survey, it will go through a period of ``Survey Validation'' (SV). During this time, the scientific capabilities of the instrument will be assessed, and decisions will be made about observational strategy during the main survey. It will involve the measurement of approximately 50,000--100,000 QSO target spectra, which will be classified via an extensive VI effort. During this phase \RR{} will be run on all spectra, and comparison with VI results will be helpful for diagnosing and addressing any common failure modes. However, at this stage we will not yet be able to train new \QN{} or \SQ{} models on DESI data, and so their use will be limited. As both (e)BOSS and DESI measure spectra over a similar wavelength range --- 3600--10000~\AA{} for (e)BOSS, and 3600--9800\AA{} for DESI --- it is possible that models trained on BOSS data will perform well when classifying DESI data. If so, our set of automatic classifiers could be used to ``sense check'' VI: if automatic classifiers confidently disagree with a first VI result, then that spectrum could be flagged to go through another round of VI. This would provide a useful aide to the VI effort, and could also prove instructive in understanding the classifiers themselves.

Once SV has been completed, we will be able to train new \QN{} and \SQ{} models on DESI data, using the VI results from SV as a truth table. We check that these 50,000--100,000 spectra will be sufficient to train high-performing \QN{} models in \secref{subsec:Dependence of QuasarNET on training set size}. Upon entering the main survey, QSO classifiers will be required for two main purposes: selecting high-$z$ QSOs for reobservation, and constructing QSO catalogues. These are discussed in \secref{subsec:Selecting high-z QSOs for reobservation} and \secref{subsec:Constructing a QSO catalogue} respectively. We assess our classifiers' performance at these tasks using spectra from BOSS DR12Q Superset. The exact properties of these spectra will differ slightly to those from DESI for a variety of reasons. For example, DESI will observe QSO targets to approximately 1 magnitude fainter than BOSS, and will use different target selection strategies. However, due to DESI's larger telescope, the per-pixel signal-to-noise ratio (SNR) for the faintest objects in BOSS and DESI will be similar, and thus BOSS DR12Q Superset data can be considered a reasonable approximation to DESI data when testing QSO selection strategies.

\subsection{Selecting high-$z$ QSOs for reobservation}
\label{subsec:Selecting high-z QSOs for reobservation}

Whereas all QSO targets in BOSS/eBOSS were allocated four, consecutively-taken, 900~second exposures on the same night, DESI will only carry out four observations (also of 900 seconds each) of spectra considered likely to be high-$z$ ($z\geq2.1$) QSOs, and these will be distributed through the time period of the survey. Such objects benefit significantly from additional exposures as they are used for \Lya{} forest analyses, which use the values of individual pixels in each QSO's spectrum and so are directly sensitive to its signal-to-noise ratio (SNR). Other objects --- contaminants and low-$z$ QSOs --- do not benefit from additional exposures provided classifications and redshifts can be accurately determined, and so will not generally be reobserved. This change allows for greater efficiency in DESI's data collecting, prioritising observations which will provide greatest scientific yield. However, it also introduces new challenges; we will need to be able to select high-$z$ QSOs for reobservation from their first exposures, which will have lower SNR than the final, coadded spectra. This reduction in SNR will make spectra harder to classify, necessitating careful thought about optimal selection methods. Further, the process of selecting for reobservation will need to be conducted entirely automatically, and should be built into the DESI pipeline in order to maximise efficiency and reducing the possibility of human error in generating selections. As such, no VI will be involved and we will need to rely entirely on automatic classifiers.

Ahead of DESI, we would like to assess which selection strategies perform best in this context. In order to approximately replicate the reduction in SNR, we construct a single-exposure dataset from BOSS DR12Q Superset data, choosing one exposure at random from each set of exposures that were coadded to make BOSS' final spectra. This ensures that our single-exposure dataset contains the same number of spectra and balance of contaminants as the coadded dataset. Using this single-exposure dataset, we obtain classifications from \RR{} and \QN{} from which we can build selection strategies. We run \RR{} using the \texttt{andmask} option, which sets pixels' inverse variance to zero if they were masked during the BOSS data reduction procedure (due to the presence of strong sky lines, for example). We then train a \QN{} model on 10\% of spectra in our single-exposure dataset, a realistic training set size that we can expect from DESI SV (see \secref{subsec:Dependence of QuasarNET on training set size} and \secref{subsec:Dependence of QuasarNET on SNR} for analyses of the effects of size and SNR of \QN{} training sets). We apply this model to all remaining single-exposure spectra from objects that were not included in the training set. We restrict our \RR{} classifications to this same set of spectra to ensure consistency.

In order to assess different classification strategies, we must consider what metrics are most relevant for the problem at hand. As a primary concern, any selection procedure must ensure that as great a proportion of true high-$z$ QSOs as possible are chosen to be reobserved. As such, we certainly would like to measure the fraction of high-$z$ QSOs that are selected by a strategy, a quantity related closely to completeness (see \eqref{eqn:completeness} and surrounding discussion). However, unlike when computing completeness, we are now not interested in whether a strategy is able to determine an accurate redshift. Rather, we only care that a strategy can correctly identify whether an object is a QSO and whether it is at high $z$; we can consider a classification as correct provided these two criteria alone are met. Of course, any strategy must also avoid selecting too many contaminants for reobservation in order to ensure efficient use of available fibres. This is best assessed by calculating the total number density of objects selected rather than the purity, as it allows us to determine directly whether a strategy is recommending a number density of objects for reobservation that is feasible for DESI. Within current DESI plans, the number density of fibres that will be assigned to reobserving high-$z$ QSOs will be approximately 50~\psqd{}. There will a limited degree of flexibility around this number, however a strategy that selects a substantially higher number density of objects than this will not be feasible, and so can be discarded.

\begin{figure}
\centering
\includegraphics[width=15cm]{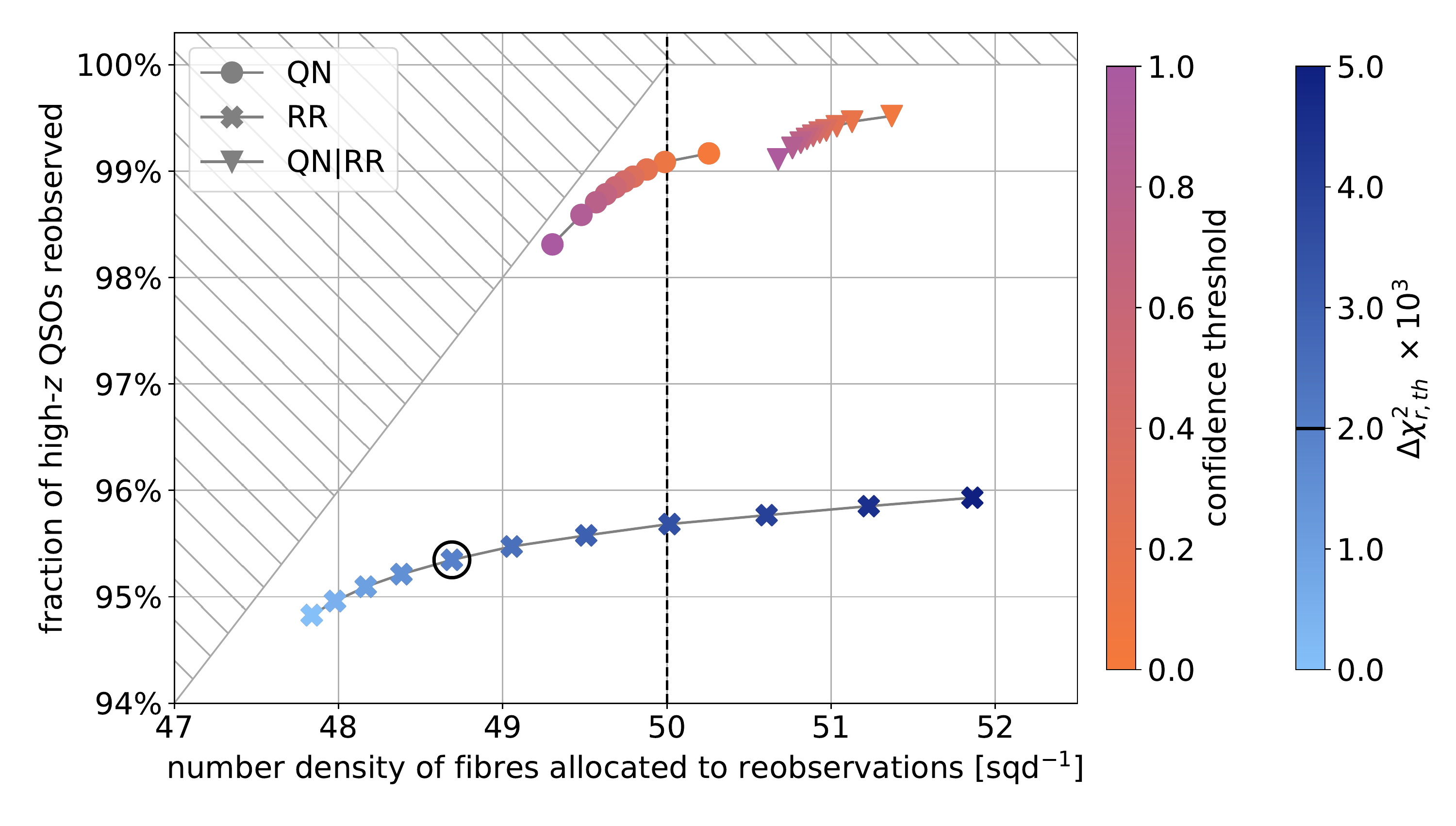}
\caption{Performance of different strategies when selecting high-$z$ QSOs for reobservation, using randomly-chosen single exposures from BOSS DR12Q Superset data. The $x$-axis has been normalised here to assume a fiducial true number density of high-$z$ QSOs of 50~\psqd{}, as is realistic for DESI (indicated by the vertical, dashed, black line). Here, the ``QN'' strategy selects objects for which \QN{} identifies at least one emission line with confidence exceeding a confidence threshold, while the ``RR'' strategy selects objects for which \RR{} finds a high-$z$ QSO template with $\Delta\chi_r^2<\Delta\chi_{r,th}^2$, as defined in \eqref{eqn:Reduced chi-squared threshold}. In the ``QN'' and ``RR'' strategies, the variation in the colour of the points indicates the choice of confidence threshold and $\Delta\chi_{r,th}^2$ respectively. In the ``QN$|$RR'' strategy, we select objects identified as high-$z$ QSOs either by \QN{} \textit{or} \RR{}. Here, we fix $\Delta\chi_{r,th}^2=2\times10^{-3}$ --- the corresponding point in RR is circled --- and variation in the colour of the points once again indicates the choice of \QN{} confidence threshold.}
\label{fig:classifier_performance_selecting_for_reobservation_randexp_simple_2cb}
\end{figure}

In \figref{fig:classifier_performance_selecting_for_reobservation_randexp_simple_2cb}, we place a number of strategies in the plane defined by these two quantities: the fraction of high-$z$ QSOs selected on the $y$-axis, and the number density of fibres assigned to reobservations on the $x$-axis. Strategies can be defined by the results of a single classifier, or by combination of results from more than one classifier. The number density of fibres allocated to reobservations of potential high-$z$ QSOs is normalised by assuming a true number density of high-$z$ QSOs of 50~\psqd{}, as indicated by the vertical, dashed line. As such, an ideal strategy would sit at the point $(50,100\%)$: it would select 100\% of high-$z$ QSOs, while only recommending a number density of reobservations of 50~\psqd{} (i.e. including no contaminants). Including some contaminants in the set of objects recommended for reobservation would then move this strategy to the right, while missing high-$z$ QSOs without including contaminants would move it diagonally downwards and to the left, along the grey line. The hashed region in the upper left corner of the plot denotes a section of the plane in which a strategy cannot sit: at a fixed number density of reobservations $N_\mathrm{reobs}\leq50$, the maximum fraction of high-$z$ QSOs reobserved is $N_\mathrm{reobs}/50$ (achieved when no contaminants are included). A realistic strategy includes both some degree of contamination and fails to select some high-$z$ QSOs, and thus sits in the non-hashed region of the plane.

\subsubsection{Performance of individual classifiers}
\label{subsubsec:Performance of individual classifiers (reobs)}

In \figref{fig:classifier_performance_selecting_for_reobservation_randexp_simple_2cb}, the results from a strategy using \RR{} classifications alone are represented by the set of blue crosses (labelled ``RR''). The most straightforward strategy that could be defined from our \RR{} results would select all spectra for which \RR{} returned a high-$z$ QSO as its best-fit template (i.e. that with the lowest $\chi^2$). However, we find that we are able to improve performance by making two simple changes to this strategy. First, we select based on the reduced chi-squared value $\chi_r^2=\chi^2/\nu$ where $\nu=n_d-n_p$ is the number of degrees of freedom, defined as the difference between the number of data points $n_d$ and the number of fit parameters $n_p$. This provides a more fair comparison between different spectral types which have different values of $n_p$. Second, we do not select based solely on the template with the lowest $\chi_r^2$. Instead, we consider all templates whose $\chi_r^2$ values fall close to the minimum, selecting an object for reobservation if at least one such template corresponds to a high-$z$ QSO. Formally, we consider templates with
\begin{equation}
\label{eqn:Reduced chi-squared threshold}
    \chi_r^2-\chi_{r,min}^2=\Delta\chi_r^2<\Delta\chi_{r,th}^2,
\end{equation}
where $\chi_{r,min}^2$ is the minimum $\chi_{r}^2$ over all template-redshift combinations, and $\Delta\chi_{r,th}^2$ is a threshold value that can be chosen freely. We present results for a range of threshold values in \figref{fig:classifier_performance_selecting_for_reobservation_randexp_simple_2cb}, from $\Delta\chi_{r,th}^2=0$ to $5\times10^{-3}$. The value of $\Delta\chi_{r,th}^2$ at each point is indicated by the corresponding colour bar to the right of the main panel. The circled point corresponds to a choice of $\Delta\chi_{r,th}^2=2\times 10^{-3}$, and is the choice of threshold used when constructing the ``QN$|$RR'' strategy presented in \secref{subsubsec:Performance of combined strategies (reobs)}. One may also consider using \RR{}'s \texttt{archetypes} option. This compares stellar and galaxy template fits to a suite of archetypes, penalising fits which show non-physical features such as negative emission lines. This reduces the number of low-$\chi_r^2$ stellar and galaxy fits, resulting in more objects being classified as high-$z$ QSOs. Subsequently, using the \texttt{archetypes} option selects a greater fraction of high-$z$ QSOs as well as a greater number of contaminants, requiring 49.8 fibres~\psqd{} to select 96.6\% of high-$z$ QSOs. If using \RR{} alone, this may be deemed a better performing strategy than the RR points shown in \figref{fig:classifier_performance_selecting_for_reobservation_randexp_simple_2cb}. However, when combined with results from \QN{} (see \secref{subsubsec:Performance of combined strategies (reobs)}), using the \texttt{archetypes} option results in a greater number density of fibres being assigned to reobservations with negligible gain in the fraction of high-$z$ QSOs selected. As such, we do not present these results here for clarity.

The results from a strategy using \QN{} classifications alone are represented by the sequence of circular points and the line that joins them (labelled ``QN''). Here, we select a spectrum if \QN{} detects at least one emission line with confidence $c>c_\mathrm{th}$, where $c_\mathrm{th}$ is the confidence threshold (discussed in \secref{subsubsec:Machine learning classifiers}). The colours of the QN points correspond to the choice of this confidence threshold (as indicated by the colour bar to the right of the main panel), with values varying from 0.05 to 0.95 in \figref{fig:classifier_performance_selecting_for_reobservation_randexp_simple_2cb}. The value of $c_\mathrm{th}$ is free to be chosen: reducing $c_\mathrm{th}$ produces a less stringent selection strategy, resulting in a greater fraction of high-$z$ QSOs being selected, as well as a greater number density of fibres being allocated to reobservations. 

The QN strategy performs better than RR when selecting QSOs for reobservation. The proportion of high-$z$ QSOs missed decreases from 4.1--5.2\% for RR, to 0.9--1.7\% for QN. For the threshold values shown, the number density of fibres allocated to reobservations depends varies within 47.8--51.8~\psqd{} for RR and 49.3--50.3~\psqd{} for QN, all of which are feasible values for DESI. As described, the values of $\Delta\chi_{r,th}^2$ and $c_\mathrm{th}$ can be varied to alter the properties of these two strategies. Indeed, values could be chosen automatically to suit the number of fibres available at any given sky location and on any given night. 

\subsubsection{Performance of combined strategies}
\label{subsubsec:Performance of combined strategies (reobs)}

We also define two simple strategies to combine classifications from \QN{} and \RR{}. First, we define a strategy ``QN\&RR'', which selects an object to be reobserved only if both the QN and RR strategies do so. This strategy represents a more stringent selection criteria than either the QN or RR strategies, and so selects fewer high-$z$ QSOS as well as reducing the number density of fibres allocated to reobservations. In the context of selecting high-$z$ QSOs for reobservation, we would like to prioritise increasing the fraction of high-$z$ QSOs reobserved over reducing the number of contaminants reobserved (within reason). As such, the QN\&RR strategy is not preferable to QN or RR as it results in at least 5.0\% of high-$z$ QSOs being missed and so we do not present this option in \figref{fig:classifier_performance_selecting_for_reobservation_randexp_simple_2cb}.

Next, we define a strategy ``QN$|$RR'', which selects an object to be reobserved if either the QN or the RR strategies does so. The results for this strategy are represented in \figref{fig:classifier_performance_selecting_for_reobservation_randexp_simple_2cb} by the coloured downwards-pointing triangles and their adjoining line. These points sit further to the right and higher in the plane than either the QN or the RR strategies, reflecting the less stringent selection criteria of the QN$|$RR strategy and the greater number of objects selected. Combining \QN{} and \RR{}'s classifications in this way results in a reduced proportion of high-$z$ QSOs that are missed, with only 0.5--0.9\% not selected. At the same time, it does not require an excessive number of fibres be allocated to reobservations: between 50.6 and 51.3~\psqd{} for different choices of \QN{}'s confidence threshold. 

As such, provided that 50.6--51.3 fibres~\psqd{} can be assigned to potential high-$z$ QSOs, adopting a QN$|$RR strategy would appear to be the ideal strategy of those presented in \figref{fig:classifier_performance_selecting_for_reobservation_randexp_simple_2cb}. However, the gain in the fraction of high-$z$ QSOs selected over the QN strategy is not dramatic. Indeed, if fewer fibres are available for reobservations, then the QN strategy provides a suitable, high-performing alternative. These results demonstrate \QN{} alone is able to provide a high-performing selection strategy, while combining it with results from \RR{} in a QN$|$RR strategy can boost performance further still.

\subsection{Constructing a QSO catalogue}
\label{subsec:Constructing a QSO catalogue}

The second QSO classifier purpose during DESI's main survey is the construction of QSO catalogues. These catalogues will then be used to measure large-scale structure in two ways: via the clustering of QSOs at redshifts $0.9<z<2.1$, and via clustering of the \Lya{} forest at $z>2.1$. In both cases, it is of great importance that the QSO catalogues contain minimal levels of contamination to ensure that the data used in these analyses truly traces the matter density at the relevant redshifts. At the same time, we need to ensure that we make maximal use of DESI's observations, without discarding significant numbers of spectra unnecessarily. As such, an optimal catalogue construction method should minimise contamination as a priority, while only maximising completeness as a secondary concern. In order to improve catalogue purity and completeness, DESI might be able to include a moderate level of visual inspection, as in the DR14Q \cite{Paris:2018A&A...613A..51P} and DR16Q \cite{Lyke:2020arXiv200709001L} catalogues from eBOSS. This VI can be targeted at specific spectra which automatic classifiers failed to classify confidently, and can be built into any catalogue construction strategy according to DESI's VI capabilities. 

\begin{figure}
\centering
\includegraphics[width=15cm]{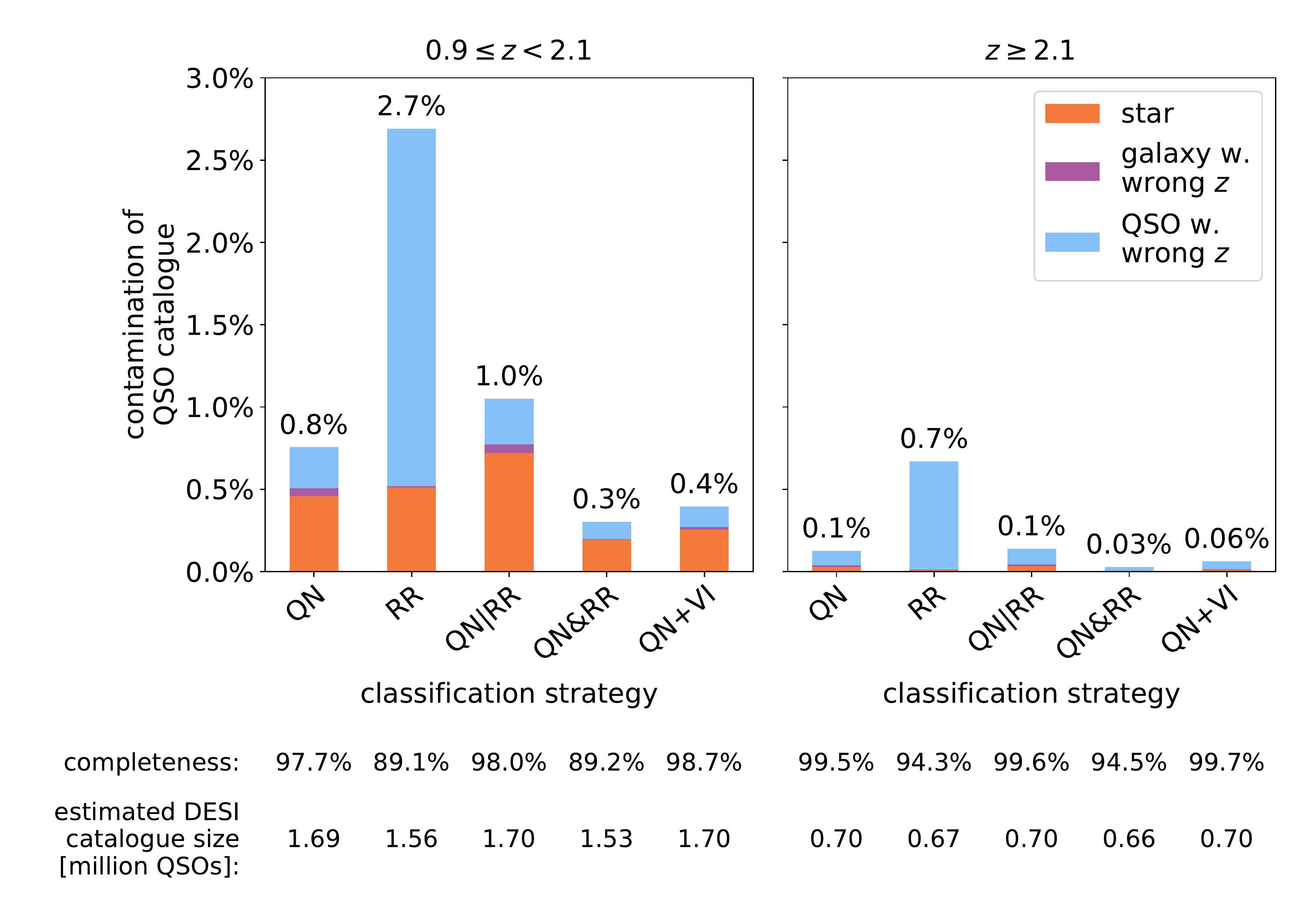}
\caption{Performance of different classification strategies when constructing a QSO catalogue. Each column shows the percentage level of contamination in the QSO catalogue, divided into different contaminating classes. In the left panel, we show performance on objects with predicted redshift $0.9\leq z<2.1$, and on the right on objects with predicted redshift $\geq2.1$. Below the panels, we also display the completeness of QSOs and an estimated DESI catalogue size for each strategy, in each $z$-bin. The estimated catalogue size is based on both the level of contamination and the level of completeness, and assumes that there will be $1.7\times10^6$ true QSOs at $z<2.1$, and $0.7\times10^6$ true QSOs at $z\geq2.1$ in DESI \cite{DESICollaboration:2016arXiv161100036D}.}
\label{fig:classifier_performance_constructing_a_catalogue_final}
\end{figure}

In order to test various strategies, we apply classifiers to coadded spectra from BOSS DR12Q Superset data. We apply \RR{} to all spectra in this dataset, using \RR{}'s \texttt{andmask} option as in \secref{subsec:Selecting high-z QSOs for reobservation}. In this case, as we are working with coadded spectra, this option sets a pixel's inverse variance to zero only when it was masked for all exposures used in constructing the coadded spectrum. When this is the case, a pixel's value cannot be considered reliable, and so should not be used in any fits. We then train a \QN{} model on 10\% of coadded spectra from DR12Q Superset, a realistic training set size to expect from DESI SV (see \secref{subsec:Dependence of QuasarNET on training set size}). We apply this model to all remaining coadded spectra from objects that were not included in the training set. Once again, we restrict our \RR{} classifications to this same set of spectra to ensure consistency.

In \figref{fig:classifier_performance_constructing_a_catalogue_final} we show the level of contamination of QSO catalogues constructed using various strategies. Results are split into low ($0.9<z<2.1$) and high ($z\geq2.1$) redshift bins in order to reflect the two separate uses of QSO catalogues in DESI, and to highlight differences in performance for each case. For each strategy, the level of contamination is broken down into three subsections: contamination by stars, by galaxies with incorrect redshifts\footnote{As mentioned in \secref{subsubsec:Performance of classifiers}, the classification of a galaxy spectrum as belonging to a QSO is considered correct provided the classifier is able to determine the correct redshift.}, and by QSOs with incorrect redshifts. As in \secref{subsubsec:Performance of classifiers}, we define a given redshift as correct if implies a velocity error $\mathrm{d}v\leq 6000$~\kms{}, ruling out catastrophic failures without requiring a high precision measurement. Below each column, we also present the level of completeness achieved by each strategy, as well as combining contamination and completeness to estimate the size of a final DESI catalogue that each strategy would yield. This catalogue size is constructed by assuming that there will be $1.7\times10^6$ true QSOs at $0.9<z<2.1$, and $0.7\times10^6$ true QSOs at $z\geq2.1$ amongst DESI's QSO targets \cite{DESICollaboration:2016arXiv161100036D}.

\subsubsection{Performance of individual classifiers}
\label{subsubsec:Performance of individual classifiers (cat)}

We first construct strategies using the classifiers \RR{} and \QN{} individually. We define a strategy ``RR'' using only results from \RR{}. Unlike in \secref{subsec:Selecting high-z QSOs for reobservation}, we are now seeking to minimise contamination as a priority rather than maximising completeness, and so we do not consider template fits other than that with the lowest $\chi^2$ (\RR{}'s best fit). However, the ``RR'' strategy presented here does incorporate \RR{}'s \texttt{ZWARN} flags --- if a spectrum raises such a flag, it is deemed not to yield a reliable classification and so is discarded. This strategy results in 2.7\% contamination and 89.1\% completeness at low $z$, and 0.7\% contamination and 94.3\% completeness at high $z$. At both low and high $z$ the majority of contaminants are QSOs with incorrect redshifts, with a moderate number of stars at low $z$ as well. As in \secref{subsubsec:Performance of individual classifiers (reobs)}, we could use \RR{}'s \texttt{archetypes} option when constructing the RR strategy in \figref{fig:classifier_performance_constructing_a_catalogue_final}. At low $z$, this results in a substantial increase in contamination to 4.8\% (from 2.7\%), while completeness increases to 94.3\% (from 89.1\%). At high $z$, there is no change in the contamination level (though the proportion of stellar contaminants increases slightly), and completeness increases to 95.8\% (from 94.3\%). As such, at low $z$ using the \texttt{archetypes} option produces a substantially less effective strategy, whereas at high $z$ it increases performance slightly. As we would like to prioritise low contamination when constructing QSO catalogues, we continue to present results without using the \texttt{archetypes} option in \figref{fig:classifier_performance_constructing_a_catalogue_final}.

We then define a strategy using results from \QN{} only, labelled ``QN''. Here, we include an object in our catalogue if \QN{} is able to identify at least one emission line in the object's spectrum with confidence $c>c_\mathrm{th}=0.5$. This value could of course be varied, but we choose to present a single value for clarity. Using this value of $c_\mathrm{th}$ results in 0.8\% contamination at low $z$, while achieving 97.7\% completeness. Here, over half of the contaminants correspond to stars, while the remainder are mostly QSOs with incorrect redshifts. At high $z$ the level of contamination drops to 0.1\% while completeness rises to 99.5\%. Contaminants here are mostly made up of QSOs with incorrect redshifts, with a small number of stars as well. It is also possible to classify using $n_\mathrm{detect}=2$, i.e. requiring that \QN{} must find at least \textit{two} emission lines with confidence greater than $c_\mathrm{th}$. For a fixed choice of $c_\mathrm{th}$, this results in a more stringent classification criteria, yielding less contamination and reducing the level of completeness as well. Again using $c_\mathrm{th}=0.5$, at low $z$ this ``2-line'' strategy achieves 0.5\% contamination at 96.9\% completeness, whilst at high $z$ it achieves 0.09\% contamination at 99.4\% completeness. We can also require that the two emission lines identified have consistent redshifts, discarding spectra for which \QN{} confidently identifies two lines which would imply drastically different QSO redshifts. This can further reduce contamination in both high- and low-$z$ catalogues, though the effects are small. Indeed, any number of more advanced interpretations of \QN{}'s outputs are possible, and could be motivated by the properties of the dataset being classified.

These results show a significantly higher level of contamination and lower levels of completeness in catalogues produced by the RR strategy compared to those from QN, both in the low- and the high-$z$ bins. In particular, there is much less contamination from QSOs with incorrect redshifts in the QN strategy, indicating that \QN{} achieves far fewer catastrophic redshift failures than \RR{}. However, unlike \RR{}, \QN{} is not designed to be a precise redshift fitter, achieving a spread in velocity errors of 793~\kms{} on correctly classified QSOs (see \secref{subsec:Dependence of QuasarNET on training set size}). At low-$z$, DESI QSOs must have a mean redshift error of at most 750~\kms{} \cite{DESI:2013internal}, and so \QN{} redshifts do not appear be sufficiently precise to meet this requirement. As such, a natural extension of the strategies presented here would be to use \RR{} as a post-classification redshift fitter, with \QN{}'s redshifts acting as a prior to guide \RR{} and perhaps using VI to resolve any objects for which \RR{} disagrees with \QN{}. At the moment, the use of priors is not possible within \RR{}, and of course any procedures would need to be tested on DESI data to ensure that DESI's redshift requirements are met.

While the QN strategy includes fewer QSOs with incorrect redshifts, it does introduce a greater level of contamination from galaxies with incorrect redshifts, and from stars in the high-$z$ bin. This difference in the breakdown of contaminants suggests that the two classifiers may be able to act in a complementary manner when combined, as discussed in \secref{subsubsec:Performance of combined strategies (cat)}. We can also see a substantial difference between the results in the low- and high-$z$ bins: both classifiers perform significantly better at high redshift. This is certainly to be expected of the RR strategy as \RR's templates are built from a high-$z$-dominated set of QSO spectra (see \secref{subsubsec:Template-fitting classifiers}), an issue which is currently being addressed by the developers. However this $z$-dependence is also present in the QN strategy, most likely due to the presence of the \Lya{} emission line in high-$z$ QSO spectra, a strong and distinctive feature that makes classification more straightforward.

\subsubsection{Performance of combined strategies}
\label{subsubsec:Performance of combined strategies (cat)}

As in \secref{subsec:Selecting high-z QSOs for reobservation}, we also consider two strategies to combine outputs from \RR{} and \QN{}. We define a ``QN$|$RR'' strategy which classifies a spectrum as a QSO if either \QN{} has confidence $c>0.5$ that it is a QSO, or \RR{} finds a QSO template as the best fit without any \texttt{ZWARN} flags. This strategy uses \QN{}'s redshift if \QN{} classifies as a QSO, and \RR{}'s otherwise. This redshift choice was found to be more effective than the alternative of using \RR{}'s redshift as default, and only using \QN{}'s redshift where \QN{} classified as a QSO but \RR{} did not. The QN$|$RR strategy introduces a greater degree of contamination to the catalogue than the QN strategy at low $z$, but less than the RR strategy. This counterintuitive behaviour is due to the choice of redshift mentioned previously: some spectra which were correctly classified as QSOs but assigned an incorrect redshift by the RR strategy are now assigned a correct redshift from \QN{} instead. At high $z$, the QN$|$RR strategy results in the same level of contamination as the QN strategy, and lower than the RR strategy. At the same time, it is able to achieve higher level of completeness than either the RR or the QN strategies at both low and high $z$. The gain over the QN strategy in this respect is small, however. As we wish to prioritise low contamination in our QSO catalogues, it seems that the QN$|$RR strategy is less effective than then QN strategy at low $z$, and thus would not be an advisable strategy. At high $z$, it performs negligibly better than the QN strategy.

Similarly, we define a ``QN\&RR'' strategy which classifies a spectrum as a QSO only if both \QN{} has confidence $c>0.5$ that it is a QSO, and \RR{} finds a QSO template as the best fit without any \texttt{ZWARN} flags. We also require that \QN{} and \RR{} agree on the object's redshift to within 6,000~\kms. Again, this strategy uses \QN{}'s redshift for all objects classified as QSOs. As one would expect, this strategy achieves lower contamination levels than either then QN or RR strategies, reducing it to 0.3\% at low $z$ and 0.03\% at high $z$. The completeness of the sample is also reduced relative to the individual classifier strategies, and as a result the predicted catalogue sizes are significantly lower than the true values. The very low levels of contamination achieved by the QN\&RR strategy make it well suited for constructing QSO catalogues. Of course, it results in significant drop in completeness relative to the QN strategy, but this may be deemed a necessary sacrifice in order to construct a maximally pure catalogue.

It is also worth noting that both combined strategies perform significantly better at high $z$ than low, again perhaps due to the presence of the distinctive \Lya{} emission line at $z\geq 1.92$. This suggests that there may be benefit to considering using different classification strategies when constructing catalogues for different purposes. For example, when constructing a low-$z$ QSO catalogue for QSO clustering analysis, one may judge that only the QN\&RR strategy is able to construct a sufficiently pure catalogue, despite its modest level of completeness. At high $z$, the levels of contamination in either the QN or QN\&RR strategies may be considered to be acceptable, and so the QN strategy may be favoured in order to make use of the 4.6 percentage point increase in completeness that this strategy offers. When measuring the clustering of QSOs at low $z$, however, understanding the completeness of the QSO sample more deeply is of great importance. The inclusion of any given QSO in a QSO catalogue depends on a number of other quantities both physical and observational, such as the photometric properties of the QSO and the position in the focal plane of the fibre used to observe it (see \cite{Reid:2016MNRAS.455.1553R,Laurent:2017JCAP...07..017L,Ata:2018MNRAS.473.4773A} for further discussion). As such, we emphasise that improvements to completeness must be complemented by future, detailed modelling of the entire pipeline when performing any clustering analysis.

\subsubsection{Role of visual inspection in constructing QSO catalogues}
\label{subsubsec:Role of visual inspection in constructing QSO catalogues}

As mentioned previously, DESI will have some capacity for VI when constructing QSO catalogues, which can be targeted towards spectra for which automatic classifiers were not able to yield a confident classification. In constructing the eBOSS DR14Q catalogue, approximately 3.7\% of new spectra were visually inspected in this way, chosen via a simple decision tree (see section 3.2 of ref. \cite{Paris:2018A&A...613A..51P}), while in constructing the DR16Q catalogue this dropped to approximately 2.9\% \cite{Lyke:2020arXiv200709001L}. Given the increase in data quantity that DESI will provide, it is likely that VI levels will need to be reduced further, perhaps to $\lesssim1\%$. The results of \secref{subsubsec:Performance of individual classifiers (cat)} and \secref{subsubsec:Performance of combined strategies (cat)} do not allow for any VI capacity, and could undoubtedly be improved by introducing a system to allow for spectra to be flagged for VI. There are several ways that this could be incorporated into any given strategy, and here we discuss two simple options.

Combining \QN{} with VI, one could define two confidence thresholds $c_\mathrm{th}^\mathrm{lo}$ and $c_\mathrm{th}^\mathrm{hi}$, and classify each spectrum with confidence $c$ as follows:
\begin{itemize}
    \item $c\geq c_\mathrm{th}^\mathrm{hi}$: \QN{} is sure that this spectrum is a QSO, thus classify as a QSO.
    \item $c_\mathrm{th}^\mathrm{lo}\leq c<c_\mathrm{th}^\mathrm{hi}$: \QN{} unsure whether this spectrum is a QSO, thus send it to VI.
    \item $c<c_\mathrm{th}^\mathrm{lo}$: \QN{} is sure that this spectrum is not a QSO, thus classify as a non-QSO.
\end{itemize}
This sends to VI those spectra which \QN{} is least able to classify definitively. Adopting this classification strategy with values $c_\mathrm{th}^\mathrm{lo}=0.04$ and $c_\mathrm{th}^\mathrm{hi}=0.96$ results in VI being requested for 1.0\% of spectra, an appropriate proportion for DESI. We can estimate its performance by simply assigning DR12Q Superset VI results to spectra that fall into this category; doing so yields the results labelled ``QN+VI'' in \figref{fig:classifier_performance_constructing_a_catalogue_final}. When compared with the QN strategy from the same figure, these results show a reduction in catalogue contamination levels by almost 50\% in both redshift bins, alongside an increase in completeness of one percentage point at low $z$. This combination of \QN{} and VI can thus be deemed highly effective, providing contamination rates in between those of the QN and QN\&RR strategies, while exceeding the completeness of either strategy.

Equally, one could combine VI with \RR{}'s classification results. For example, one could send all spectra with \texttt{ZWARN} flags corresponding to fitting issues\footnote{Some \texttt{ZWARN} flags correspond to issues such as broken fibres, and so VI would not be able to help with these issues.} to VI. Compared to the RR strategy, this results in substantial improvements to completeness at both low and high $z$ (3.7 and 1.6 percentage points respectively), and only small ($<0.1$ percentage point) improvements to contamination. However, including VI in this way requires 6.7\% of spectra to be visually inspected, a proportion that lies beyond the bounds of feasibility for DESI. Restricting VI to those spectra for which a \texttt{ZWARN} flag was raised and the best fit spectral type was ``QSO'', we require a far more manageable 0.6\% of spectra to be inspected. Here, there is negligible reduction in contamination, but completeness increases by at least 1 percentage point in both low- and high-$z$ bins.

Evidently, VI can also be built into more complex classification strategies as well. While in BOSS, all QSO target spectra were visually inspected, in eBOSS, decision trees were built to combine automatic classifier results with VI. In eBOSS DR14Q \cite{Paris:2018A&A...613A..51P}, a decision tree was built based on the five templates from the pipeline (\texttt{idlspec2d}) with lowest reduced $\chi^2$ values, highlighting spectra as requiring expert VI if the top 5 best-fit solutions were inconsistent, or if flags denoting low spectral quality were raised (see section 3 of ref. \cite{Paris:2018A&A...613A..51P} for further details). In DR16Q \cite{Lyke:2020arXiv200709001L}, a similar decision tree was used, with \QN{} employed to reduce the VI proportion further still. As a result of these decision trees, only 3.7\% and 2.9\% of new spectra were visually inspected in the construction of DR14Q and DR16Q respectively. These decision trees were built with prior knowledge of \texttt{idlspec2d}'s failure modes and the distribution of contaminant spectra. For example, applying \texttt{idlspec2d} to QSO target spectra from BOSS introduced a significant degree of contamination from stars, particularly at low $z$. As such, the decision tree was designed to carefully remove stellar spectra for which \texttt{idlspec2d} incorrectly returned ``QSO'' as its best fit. When constructing catalogues in DESI, a similar decision tree will be useful, constructed to take into account both the properties of \RR{}'s classifications and the contaminants in the DESI QSO target set. Such a decision tree could be constructed manually (as in eBOSS), or a simple machine learning approach could be used to assess in greater detail the full set of template fits from \RR{} and line identifications from \QN{}.
\section{Summary \& conclusions}
\label{sec:Summary conclusions}

In this work, we have assessed problems of QSO classification relevant for future spectroscopic surveys such as DESI, and have demonstrated that existing automatic classifiers can be used to construct highly effective classification strategies. In \secref{sec:Data and tools}, we summarised the automatic classifiers currently available. We provided a simple comparison of their performance levels, demonstrating that \QN{} is able to out-perform other classifiers over a range of redshifts. In \secref{sec:QSO classification in DESI}, we identified the QSO classification tasks that will be relevant to DESI's main survey: selecting high-$z$ QSOs for reobservation, and constructing QSO catalogues. We quantified how well \QN{} and \RR{} perform at addressing these tasks by applying them to BOSS DR12Q Superset data, using \QN{} models trained on appropriately-sized training sets, and using single-exposure spectra where necessary.

We then addressed the two classification tasks in turn, first presenting the performance of various strategies when selecting high-$z$ QSOs for reobservation in \secref{subsec:Selecting high-z QSOs for reobservation}. We found that reobserving all objects selected by \QN{} alone provides an effective solution, resulting in approximately 1\% of high-$z$ QSOs being lost. This loss can be reduced further --- to 0.5\% --- by reobserving all objects selected by either \QN{} or \RR{}, provided an additional 1 fibre~\psqd{} can be made available for reobservations. In \secref{subsec:Constructing a QSO catalogue}, we then used coadded spectra to consider various classification strategies in the context of constructing QSO catalogues. We showed that using \QN{} alone for this task is able to offer sub-percent levels of contamination at both low and high $z$, simultaneously yielding high levels of completeness. We showed further that including in a catalogue only QSOs identified by both \QN{} and \RR{} reduces levels of contamination by at least a factor of 2, though a substantial reduction in completeness is an unfortunate consequence. Alternatively, we considered combining \QN{}'s outputs with a small visual inspection fraction, defining a subset of uncertain spectra using two confidence thresholds. Permitting 1\% of spectra to be manually classified in this way reduced contamination to almost half that when using \QN{} alone, improving levels of completeness at the same time.

In all, we have demonstrated that \QN{} alone is able to suitably address the QSO classification tasks of DESI, and that combining its classifications with those from \RR{} in simple ways is able to boost performance further still. The exact performance levels achieved by the strategies we define may, of course, vary when applied to DESI data. They will be affected by the distribution of contaminant types through the set of QSO target spectra, and by the instrumental properties of DESI itself. The precise impact of these differences is not yet clear, though the similar pixel-level noise in BOSS and DESI's faintest QSOs indicates that BOSS data represents a suitable and well-understood proxy to use at the current moment. Further, both \QN{} and \RR{} will evolve ahead of DESI; \RR{}'s QSO templates are actively being developed to yield improved performance, while there is potential to extract information more efficiently and more precisely from \QN{}'s raw outputs.

Before DESI's main survey begins, the performance levels of these updated classifiers on DESI data will need to be tested, making use of data taken during the survey validation phase. In conjunction with an extensive VI programme, this data will allow for new \QN{} and \SQ{} models to be trained and tested alongside \RR{}. These results will then enable the development of more advanced combined-classifier strategies, taking into account the specific details of DESI's data reduction and targeting processes. In the mean time, the results presented in this work can offer encouragement that the current range of automatic QSO classifiers are well suited to addressing the needs of forthcoming spectroscopic surveys, and can safely reduce the historical reliance on visual inspection without adversely affecting science outcomes.
\section*{Acknowledgements}

We thank Nicolas Busca and Stephen Bailey for valuable discussion around the methods and application of \QN{} and \RR{} respectively. We also thank Ignasi P\'{e}rez-R\`{a}fols for similarly valuable discussion around the methods of \SQ{}, and for providing classification results from \SQ{} on BOSS DR12Q Superset data. JF was supported by a Science and Technology Facilities Council (STFC) studentship. AFR acknowledges support by an STFC Ernest Rutherford Fellowship, grant reference ST/N003853/1, and by FSE funds trough the program Ramon y Cajal (RYC-2018-025210) of the Spanish Ministry of Science and Innovation. AP was supported by the Royal Society. AP and AFR were further supported by STFC Consolidated Grant number ST/R000476/1. This work was partially supported by the UCL Cosmoparticle Initiative.

\begin{appendix}
\section{Technical tests of \QN{}}

\subsection{Dependence of \QN{}'s performance on training set size}
\label{subsec:Dependence of QuasarNET on training set size}

\begin{figure}
\centering
\includegraphics[width=15cm]{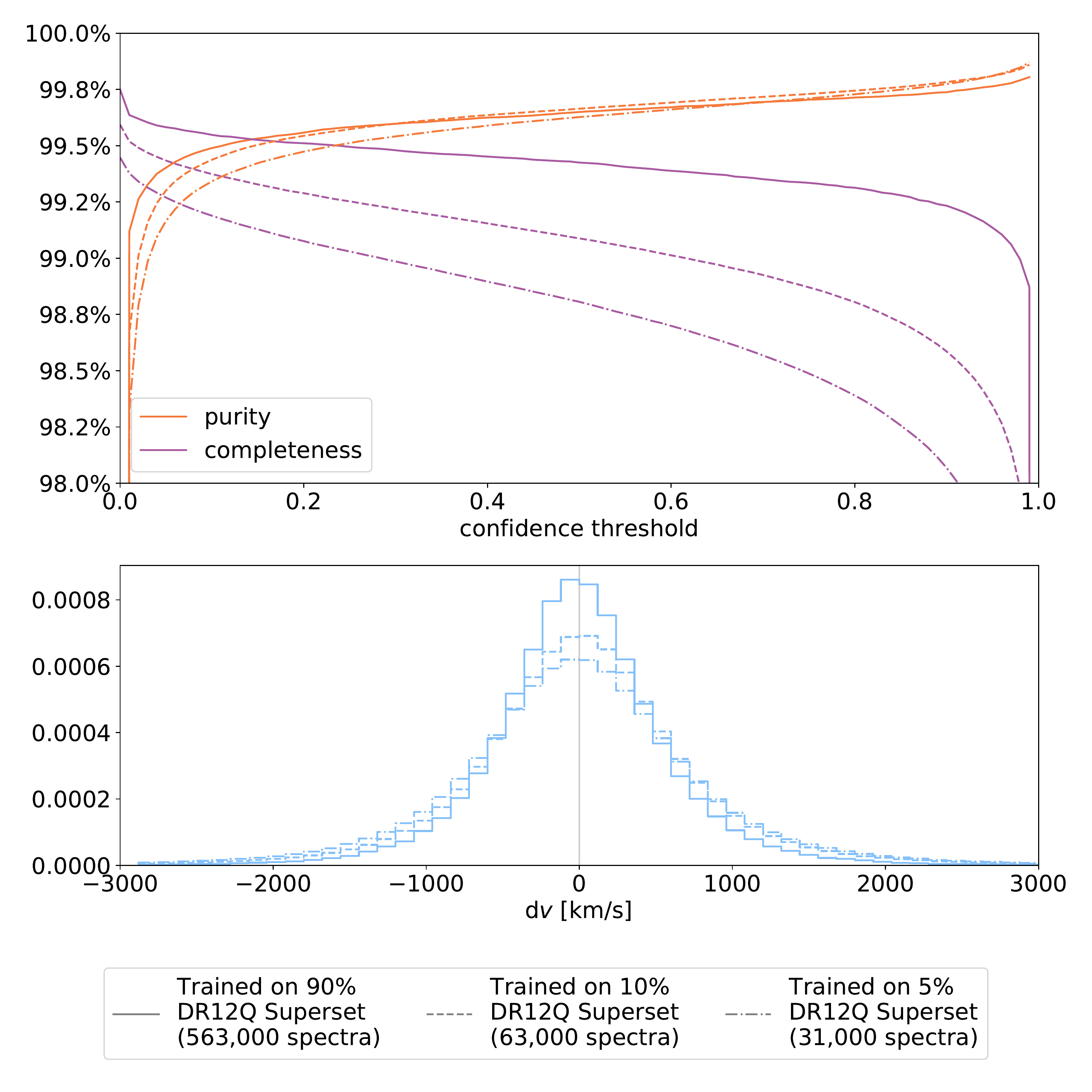}
\caption{Performance of \QN{} models with varying training set sizes. The top panel shows purity and completeness as a function of the classification confidence threshold, while the bottom panel shows a histogram of velocity errors for successful QSO classifications. Here, the solid and dashed lines --- corresponding to training sets made from 90\% and 10\% of BOSS DR12Q Superset data respectively --- show the mean performance of 10 \QN{} models. The dot-dashed line shows mean performance over 9 models with training sets made from 5\% of DR12Q Superset data, excluding one outlying model which produced degraded performance levels.}
\label{fig:qn_performance_vs_training_set_size_2panel_exclude_outliers}
\end{figure}

In ref. \cite{Busca:2018arXiv180809955B}, \QN{} models were trained on 80\% of DR12Q Superset data (approximately 500,000 spectra), while in \figref{fig:classifier_performance_summary_altcolour}, 90\% was used (approximately 560,000 spectra). These training sets are significantly larger than the $\sim$50,000--100,000 spectra that will be visually inspected during DESI's survey validation (SV) period. In order to check that this reduced number of spectra will be sufficient to train high-performing \QN{} models, we assess models trained on varying fractions of the DR12Q Superset data. We consider models trained on 10\% and 5\% of this data, approximately equivalent to $63,000$ and $31,000$ spectra respectively, and thus corresponding to ``realistic'' and ``worst case scenario'' training set sizes that we can expect from DESI SV. We compare the performances of models trained on datasets of these sizes to a ``fiducial'' training set size of 90\% of DR12 data, as was used in \figref{fig:classifier_performance_summary_altcolour}. For each training set size, we train 10 models on random subsets of DR12Q Superset data, and for each model, we test performance on spectra from all objects not in its training set. In \figref{fig:qn_performance_vs_training_set_size_2panel_exclude_outliers}, we then plot the mean performance across these 10 models for the 90\%- and 10\%-trained models, and the mean performance across 9 models for the 5\%-trained models. We exclude one outlying 5\%-trained model which showed degraded performance relative to the remaining 9 models. It is not clear why this outlier exists, and it provides motivation to aim for an SV truth table of $\gtrsim50,000$ QSO targets. If this is not feasible for any reason, then efforts should be made to investigate whether alternative \QN{} architectures --- fewer convolutional layers, for example --- are better suited to smaller training set sizes.

The upper panel of \figref{fig:qn_performance_vs_training_set_size_2panel_exclude_outliers} shows the levels of purity and completeness achieved by each model as a function of ``confidence threshold'' (as explained in \ref{subsubsec:Machine learning classifiers}). Here, we set $n_\mathrm{detect}=1$ for simplicity. Comparing the solid and dashed lines demonstrates that reducing the training set size from 90\% (solid lines) to 10\% (dashed lines) of DR12Q Superset has little effect on the model's purity, and results in only a small (0.25--0.5 percentage point) drop in the model's completeness at any given confidence threshold. The 10\%-trained model achieves ``optimal'' performance --- where purity and completeness are equal --- at a confidence threshold of 0.08, where purity and completeness are both 99.4\%. When compared to the 90\%-trained model, which achieves purity and completeness of 99.5\% at a confidence threshold of 0.15, it is clear that $\sim62,000$ training spectra is sufficient for a \QN{} model to achieve high purity and completeness in its test sample. Equally, considering the dash-dotted lines, reducing the training set size further to only 5\% of DR12Q Superset results in an additional drop in ``optimal'' performance of only 0.2 percentage points, to 99.2\% (at a confidence threshold of 0.07). Thus we can conclude that even $\sim31,000$ spectra can be sufficient to achieve high levels of purity and completeness. 

The lower panel of \figref{fig:qn_performance_vs_training_set_size_2panel_exclude_outliers} shows a (normalised) histogram of the velocity errors (relative to VI redshifts) for spectra that \QN{} correctly classified as belonging to QSOs, with velocity error less than 6000~\kms{}. Once again comparing the solid and dashed lines, we can see relatively small differences between the results obtained by the 90\%- and 10\%-trained models. While predictions from the 90\%-trained model have a median velocity error of $-8$~\kms{} and a standard deviation of $618$~\kms{}, results from the 10\%-trained model have a median of $18$~\kms{} and a standard deviation of $793$~\kms{}. As such, both models show no significant bias in the estimated redshifts, and the spread of velocity errors is ~30\% larger for the 10\%-trained model. This marks a moderate increase, though it is important to note that \QN{} is not designed to be a precision redshift fitter and so such values are not excessive. Finally, using a model trained on 5\% of DR12 data results in further spreading of the velocity error distribution, achieving a median velocity error of $-6$~\kms{}, with a standard deviation of $902$~\kms{}. Once again, there is no significant bias in the estimated redshifts for models using this training set size, and the velocity error spread does not increase to unacceptable levels. 

Clearly, when training on DESI SV data, precise performance levels of \QN{} models may vary slightly compared to those shown in \figref{fig:qn_performance_vs_training_set_size_2panel_exclude_outliers} due to, for example, differences in targeting procedures between BOSS and DESI. However, it is reasonable to expect that a \QN{} model trained on visually inspected data from DESI SV will be able to achieve similarly high levels of performance to the 10\%-trained model presented here, and as such we can be reassured that the smaller training set provided by DESI SV will not inhibit our ability to train high-performing \QN{} models. Equally, the small drop in performance when using the 5\%-trained model suggests that even in a ``worst case scenario'' in which DESI SV is impaired, effective \QN{} models would still be able to be trained.

\subsection{Dependence of \QN{}'s performance on signal-to-noise ratio}
\label{subsec:Dependence of QuasarNET on SNR}

As discussed in \secref{subsec:Selecting high-z QSOs for reobservation}, a key task for QSO classifiers in DESI will be to select high-$z$ QSOs for reobservation. This introduces a number of new challenges, most notably the need to carry out classification on spectra obtained from single exposures. These spectra will have a lower signal-to-noise ratio (SNR) than coadded spectra, making classification more difficult. Ahead of DESI, we need to test how well \QN{} models are able to classify single-exposure spectra, and also determine whether a model trained on single-exposures or coadded spectra is preferable when doing so. In order to answer these questions, we construct a single-exposure version of the DR12Q Superset data, taking spectra from one exposure chosen at random from each set of coadded exposures (excluding any low quality exposures). As such, our single-exposure dataset is of the same size as our coadded dataset, with each spectrum having a direct coadded counterpart, thus ensuring consistent balance of contaminants. Within each set of exposures used in BOSS' coadds, the BOSS pipeline identifies a ``best'' exposure, that with the highest average SNR. We could have used these best exposures rather than randomly chosen ones, but this simply improves performance of all models slightly, without affecting our qualitative conclusions.

\begin{figure}
\centering
\includegraphics[width=15cm]{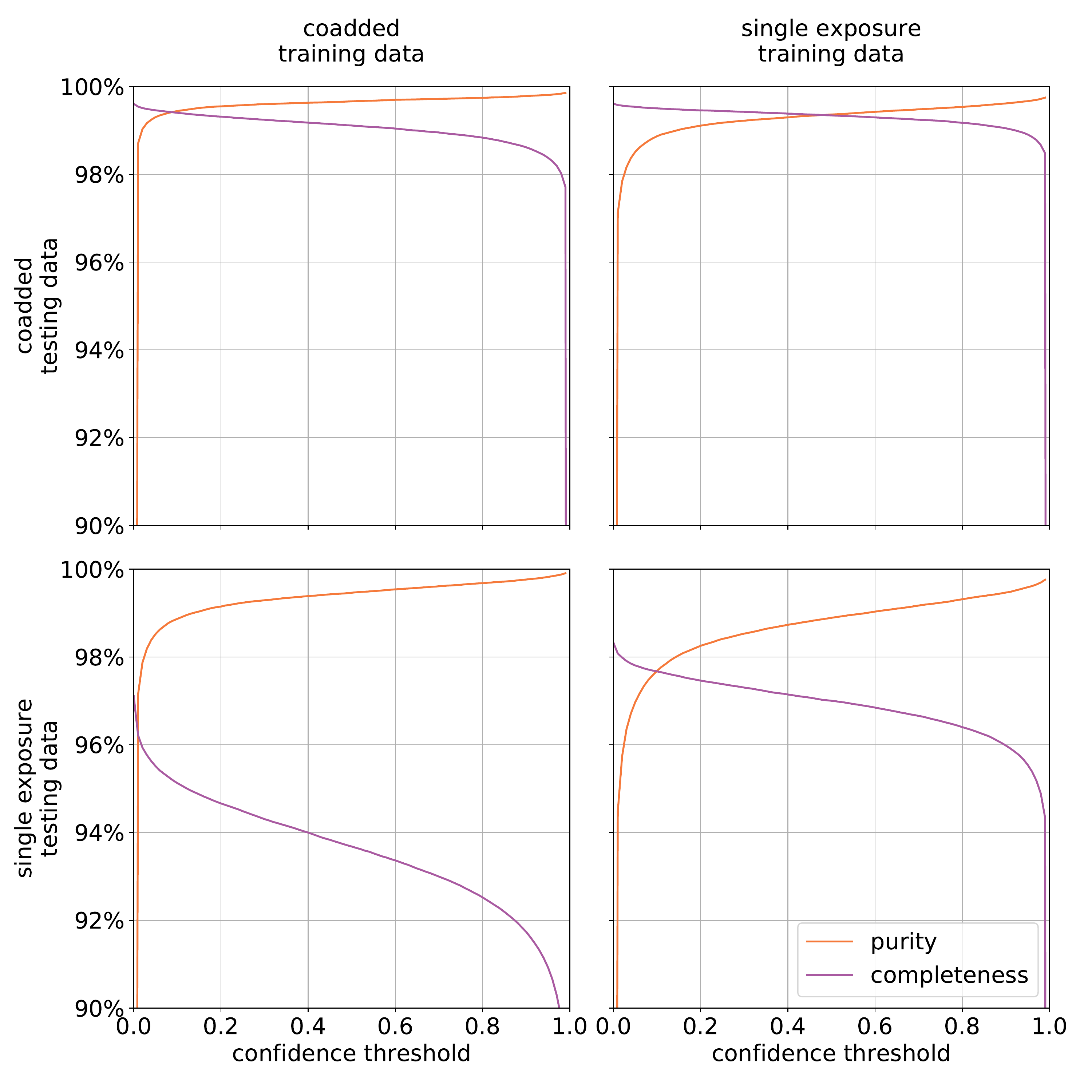}
\caption{Performance of \QN{} models with different numbers of exposures in the training and testing data. The left column corresponds to a \QN{} model trained on data consisting of coadded spectra, whereas the right column corresponds to a model trained on single-exposure spectra. Both models were trained on $\sim$10\% of DR12Q Superset spectra ($\sim62,000$ spectra), a similar quantity of visually inspected data to that which will be available from DESI SV. The top row corresponds to results obtained from applying each model to a coadded dataset, while the bottom row corresponds to a single-exposure test set. Each panel shows the purity and completeness of the classifications as a function of \QN{}'s confidence threshold.}
\label{fig:qn_performance_vs_snr}
\end{figure}

In \figref{fig:qn_performance_vs_snr}, we show the performance of \QN{} models trained and tested on coadded/single-exposure data. In the left column are results for a model trained on coadded spectra --- each made from on average 4 exposures --- from BOSS DR12. This is the same model that was used in \secref{subsec:Constructing a QSO catalogue} and \secref{subsec:Dependence of QuasarNET on training set size}, trained on 10\% of DR12Q Superset as is appropriate given the estimated size of the DESI SV dataset. In the right column are results for a model trained on single exposures from DR12, using 10\% of the single-exposure dataset for training, as was used in \secref{subsec:Selecting high-z QSOs for reobservation}. In the top row of \figref{fig:qn_performance_vs_snr} are results when models are applied to a coadded test dataset, while in the bottom row are results when applied to a single-exposure test dataset. For each model, we use a test set of DR12Q Superset spectra from all objects that were not included in that model's training set. We also exclude from our test sets spectra which were classified with anything other than maximal confidence by the DR12Q Superset VI procedure. As such, the results shown in the top left panel are equivalent to the results shown by the dashed lines in the top panel of \figref{fig:qn_performance_vs_training_set_size_2panel_exclude_outliers}.

Moving from the top left to the top right panel, we are reducing the SNR in our training data while maintaining a high SNR in our test data. This change results in a slight increase in the completeness at a given confidence threshold, but a moderate drop in the purity of $\sim$0.5 percentage points. These differences are due to the single-exposure-trained model classifying more objects as QSOs for a given confidence threshold, of which most are incorrect. Conversely, moving from the top left to the bottom left panel, we are reducing the SNR in our test data while maintaining a high SNR in our training data. This results in a substantial drop in completeness ($\sim$5--8 percentage points) due to the model correctly identifying fewer single-exposure QSO spectra for a given confidence threshold. Equally, it results in a moderate drop in purity ($\sim$1 percentage point), mostly due to incorrect redshift determination in the single-exposure QSO spectra that are classified as QSOs. Finally, moving from the bottom left to the bottom right panel, we are reducing the SNR in our training data while maintaining a low SNR in our test data. This results in a $\sim$3--4 percentage point increase in completeness as the single-exposure-trained model is better able to identify low-SNR QSO spectra, but a $\sim$0.5--1 percentage point drop in purity due to contamination both by stellar spectra and incorrectly determined redshifts. 

From these four sets of results, we can draw a number of conclusions. When classifying coadded spectra, a single-exposure-trained model is able to achieve a higher level of completeness than a coadd-trained model at a given confidence threshold. However, this comes at the expense of purity, and it is preferable to use a coadd-trained model with a lower confidence threshold if completeness is a priority. As such, a coadd-trained model is almost certainly preferable. A parallel conclusion can be drawn when classifying single-exposure spectra. A coadd-trained model is able to produce a more pure set of QSOs, but achieves substantially lower levels of completeness than a single-exposure-trained model when using the same confidence threshold. Again, it is almost always possible to match the coadd-trained model's purity with the single-exposure-trained model by simply increasing the confidence threshold, with a higher completeness being achieved at the same time. As such, it is preferable to use a single-exposure-trained model in this case.

Most importantly, we can be reassured that \QN{} is able to produce models that perform well on single-exposure data. From the bottom right panel of \figref{fig:qn_performance_vs_snr}, we can see that a single-exposure-trained model achieves purity and completeness of 97.7\% on single-exposure test data at a confidence threshold of 0.1. While there is a drop in performance compared to coadded data, this is to be expected: classifying low-SNR spectra is an inherently more difficult task. In particular, it is possible that single-exposure spectra classified incorrectly by \QN{} would not have been confidently classified by a VI expert either, but we are able to assign that single-exposure spectrum a classification as our VI results were carried out on coadded data. 
\end{appendix}

\bibliographystyle{JHEP.bst}
\bibliography{ms}

\providecommand{\href}[2]{#2}\begingroup\raggedright\begin{thebibliography}{10}

\bibitem{Ata:2018MNRAS.473.4773A}
M.~{Ata}, F.~{Baumgarten}, J.~{Bautista}, F.~{Beutler}, D.~{Bizyaev}, M.~R.
  {Blanton} et~al., \emph{{The clustering of the SDSS-IV extended Baryon
  Oscillation Spectroscopic Survey DR14 quasar sample: first measurement of
  baryon acoustic oscillations between redshift 0.8 and 2.2}},
  \href{https://doi.org/10.1093/mnras/stx2630}{\emph{\mnras} {\bfseries 473}
  (2018) 4773} [\href{https://arxiv.org/abs/1705.06373}{{\ttfamily
  1705.06373}}].

\bibitem{Hou:2020arXiv200708998H}
J.~{Hou}, A.~G. {S{\'a}nchez}, A.~J. {Ross}, A.~{Smith}, R.~{Neveux},
  J.~{Bautista} et~al., \emph{{The Completed SDSS-IV extended Baryon
  Oscillation Spectroscopic Survey: BAO and RSD measurements from anisotropic
  clustering analysis of the Quasar Sample in configuration space between
  redshift 0.8 and 2.2}}, {\emph{arXiv e-prints} (2020) arXiv:2007.08998}
  [\href{https://arxiv.org/abs/2007.08998}{{\ttfamily 2007.08998}}].

\bibitem{Neveux:2020arXiv200708999N}
R.~{Neveux}, E.~{Burtin}, A.~{de Mattia}, A.~{Smith}, A.~J. {Ross}, J.~{Hou}
  et~al., \emph{{The Completed SDSS-IV extended Baryon Oscillation
  Spectroscopic Survey: BAO and RSD measurements from the anisotropic power
  spectrum of the Quasar sample between redshift 0.8 and 2.2}}, {\emph{arXiv
  e-prints} (2020) arXiv:2007.08999}
  [\href{https://arxiv.org/abs/2007.08999}{{\ttfamily 2007.08999}}].

\bibitem{deSainteAgathe:2019A&A...629A..85D}
V.~{de Sainte Agathe}, C.~{Balland}, H.~{du Mas des Bourboux}, N.~G. {Busca},
  M.~{Blomqvist}, J.~{Guy} et~al., \emph{{Baryon acoustic oscillations at z =
  2.34 from the correlations of Ly{\ensuremath{\alpha}} absorption in eBOSS
  DR14}}, \href{https://doi.org/10.1051/0004-6361/201935638}{\emph{\aap}
  {\bfseries 629} (2019) A85}
  [\href{https://arxiv.org/abs/1904.03400}{{\ttfamily 1904.03400}}].

\bibitem{Blomqvist:2019A&A...629A..86B}
M.~{Blomqvist}, H.~{du Mas des Bourboux}, N.~G. {Busca}, V.~{de Sainte Agathe},
  J.~{Rich}, C.~{Balland} et~al., \emph{{Baryon acoustic oscillations from the
  cross-correlation of Ly{\ensuremath{\alpha}} absorption and quasars in eBOSS
  DR14}}, \href{https://doi.org/10.1051/0004-6361/201935641}{\emph{\aap}
  {\bfseries 629} (2019) A86}
  [\href{https://arxiv.org/abs/1904.03430}{{\ttfamily 1904.03430}}].

\bibitem{duMasdesBourboux:2020arXiv200708995D}
H.~{du Mas des Bourboux}, J.~{Rich}, A.~{Font-Ribera}, V.~{Sainte de Agathe},
  J.~{Farr}, T.~{Etourneau} et~al., \emph{{The Completed SDSS-IV extended
  Baryon Oscillation Spectroscopic Survey: Baryon acoustic oscillations with
  Lyman-$\alpha$ forests}}, {\emph{arXiv e-prints} (2020) arXiv:2007.08995}
  [\href{https://arxiv.org/abs/2007.08995}{{\ttfamily 2007.08995}}].

\bibitem{Schmidt:1983ApJ...269..352S}
M.~{Schmidt} and R.~F. {Green}, \emph{{Quasar evolution derived from the
  Palomar bright quasar survey and other complete quasar surveys.}},
  \href{https://doi.org/10.1086/161048}{\emph{\apj} {\bfseries 269} (1983)
  352}.

\bibitem{Hewett:1995AJ....109.1498H}
P.~C. {Hewett}, C.~B. {Foltz} and F.~H. {Chaffee}, \emph{{The Large Bright
  Quasar Survey.VI.Quasar Catalog and Survey Parameters}},
  \href{https://doi.org/10.1086/117380}{\emph{\aj} {\bfseries 109} (1995)
  1498}.

\bibitem{Croom:2004mas..conf...57C}
S.~{Croom}, B.~{Boyle}, T.~{Shanks}, P.~{Outram}, A.~{Myers}, R.~{Smith}
  et~al., \emph{{The 2DF QSO Redshift Survey}},  in \emph{Multiwavelength AGN
  Surveys}, R.~{M{\'u}jica} and R.~{Maiolino}, eds., pp.~57--62, Oct., 2004,
  \href{https://doi.org/10.1142/9789812702432_0015}{DOI}.

\bibitem{Croom:2001MNRAS.322L..29C}
S.~M. {Croom}, R.~J. {Smith}, B.~J. {Boyle}, T.~{Shanks}, N.~S. {Loaring},
  L.~{Miller} et~al., \emph{{The 2dF QSO Redshift Survey - V. The 10k
  catalogue}},
  \href{https://doi.org/10.1046/j.1365-8711.2001.04474.x}{\emph{\mnras}
  {\bfseries 322} (2001) L29}
  [\href{https://arxiv.org/abs/astro-ph/0104095}{{\ttfamily
  astro-ph/0104095}}].

\bibitem{Gunn:2006AJ....131.2332G}
J.~E. {Gunn}, W.~A. {Siegmund}, E.~J. {Mannery}, R.~E. {Owen}, C.~L. {Hull},
  R.~F. {Leger} et~al., \emph{{The 2.5 m Telescope of the Sloan Digital Sky
  Survey}}, \href{https://doi.org/10.1086/500975}{\emph{\aj} {\bfseries 131}
  (2006) 2332} [\href{https://arxiv.org/abs/astro-ph/0602326}{{\ttfamily
  astro-ph/0602326}}].

\bibitem{Schneider:2007AJ....134..102S}
D.~P. {Schneider}, P.~B. {Hall}, G.~T. {Richards}, M.~A. {Strauss}, D.~E.
  {Vanden Berk}, S.~F. {Anderson} et~al., \emph{{The Sloan Digital Sky Survey
  Quasar Catalog. IV. Fifth Data Release}},
  \href{https://doi.org/10.1086/518474}{\emph{\aj} {\bfseries 134} (2007) 102}
  [\href{https://arxiv.org/abs/0704.0806}{{\ttfamily 0704.0806}}].

\bibitem{Schneider:2010AJ....139.2360S}
D.~P. {Schneider}, G.~T. {Richards}, P.~B. {Hall}, M.~A. {Strauss}, S.~F.
  {Anderson}, T.~A. {Boroson} et~al., \emph{{The Sloan Digital Sky Survey
  Quasar Catalog. V. Seventh Data Release}},
  \href{https://doi.org/10.1088/0004-6256/139/6/2360}{\emph{\aj} {\bfseries
  139} (2010) 2360} [\href{https://arxiv.org/abs/1004.1167}{{\ttfamily
  1004.1167}}].

\bibitem{Dawson:2013AJ....145...10D}
K.~S. {Dawson}, D.~J. {Schlegel}, C.~P. {Ahn}, S.~F. {Anderson},
  {\'E}.~{Aubourg}, S.~{Bailey} et~al., \emph{{The Baryon Oscillation
  Spectroscopic Survey of SDSS-III}},
  \href{https://doi.org/10.1088/0004-6256/145/1/10}{\emph{\aj} {\bfseries 145}
  (2013) 10} [\href{https://arxiv.org/abs/1208.0022}{{\ttfamily 1208.0022}}].

\bibitem{Eisenstein:2011AJ....142...72E}
D.~J. {Eisenstein}, D.~H. {Weinberg}, E.~{Agol}, H.~{Aihara}, C.~{Allende
  Prieto}, S.~F. {Anderson} et~al., \emph{{SDSS-III: Massive Spectroscopic
  Surveys of the Distant Universe, the Milky Way, and Extra-Solar Planetary
  Systems}}, \href{https://doi.org/10.1088/0004-6256/142/3/72}{\emph{\aj}
  {\bfseries 142} (2011) 72} [\href{https://arxiv.org/abs/1101.1529}{{\ttfamily
  1101.1529}}].

\bibitem{Paris:2012A&A...548A..66P}
I.~{P{\^a}ris}, P.~{Petitjean}, {\'E}.~{Aubourg}, S.~{Bailey}, N.~P. {Ross},
  A.~D. {Myers} et~al., \emph{{The Sloan Digital Sky Survey quasar catalog:
  ninth data release}},
  \href{https://doi.org/10.1051/0004-6361/201220142}{\emph{\aap} {\bfseries
  548} (2012) A66} [\href{https://arxiv.org/abs/1210.5166}{{\ttfamily
  1210.5166}}].

\bibitem{Paris:2014A&A...563A..54P}
I.~{P{\^a}ris}, P.~{Petitjean}, {\'E}.~{Aubourg}, N.~P. {Ross}, A.~D. {Myers},
  A.~{Streblyanska} et~al., \emph{{The Sloan Digital Sky Survey quasar catalog:
  tenth data release}},
  \href{https://doi.org/10.1051/0004-6361/201322691}{\emph{\aap} {\bfseries
  563} (2014) A54} [\href{https://arxiv.org/abs/1311.4870}{{\ttfamily
  1311.4870}}].

\bibitem{Paris:2017A&A...597A..79P}
I.~{P{\^a}ris}, P.~{Petitjean}, N.~P. {Ross}, A.~D. {Myers}, {\'E}.~{Aubourg},
  A.~{Streblyanska} et~al., \emph{{The Sloan Digital Sky Survey Quasar Catalog:
  Twelfth data release}},
  \href{https://doi.org/10.1051/0004-6361/201527999}{\emph{\aap} {\bfseries
  597} (2017) A79} [\href{https://arxiv.org/abs/1608.06483}{{\ttfamily
  1608.06483}}].

\bibitem{Dawson:2016AJ....151...44D}
K.~S. {Dawson}, J.-P. {Kneib}, W.~J. {Percival}, S.~{Alam}, F.~D. {Albareti},
  S.~F. {Anderson} et~al., \emph{{The SDSS-IV Extended Baryon Oscillation
  Spectroscopic Survey: Overview and Early Data}},
  \href{https://doi.org/10.3847/0004-6256/151/2/44}{\emph{\aj} {\bfseries 151}
  (2016) 44} [\href{https://arxiv.org/abs/1508.04473}{{\ttfamily 1508.04473}}].

\bibitem{Blanton:2017AJ....154...28B}
M.~R. {Blanton}, M.~A. {Bershady}, B.~{Abolfathi}, F.~D. {Albareti},
  C.~{Allende Prieto}, A.~{Almeida} et~al., \emph{{Sloan Digital Sky Survey IV:
  Mapping the Milky Way, Nearby Galaxies, and the Distant Universe}},
  \href{https://doi.org/10.3847/1538-3881/aa7567}{\emph{\aj} {\bfseries 154}
  (2017) 28} [\href{https://arxiv.org/abs/1703.00052}{{\ttfamily 1703.00052}}].

\bibitem{Paris:2018A&A...613A..51P}
I.~{P{\^a}ris}, P.~{Petitjean}, {\'E}.~{Aubourg}, A.~D. {Myers},
  A.~{Streblyanska}, B.~W. {Lyke} et~al., \emph{{The Sloan Digital Sky Survey
  Quasar Catalog: Fourteenth data release}},
  \href{https://doi.org/10.1051/0004-6361/201732445}{\emph{\aap} {\bfseries
  613} (2018) A51} [\href{https://arxiv.org/abs/1712.05029}{{\ttfamily
  1712.05029}}].

\bibitem{Lyke:2020arXiv200709001L}
B.~W. {Lyke}, A.~N. {Higley}, J.~N. {McLane}, D.~P. {Schurhammer}, A.~D.
  {Myers}, A.~J. {Ross} et~al., \emph{{The Sloan Digital Sky Survey Quasar
  Catalog: Sixteenth Data Release}}, {\emph{arXiv e-prints} (2020)
  arXiv:2007.09001} [\href{https://arxiv.org/abs/2007.09001}{{\ttfamily
  2007.09001}}].

\bibitem{Bolton:2012AJ....144..144B}
A.~S. {Bolton}, D.~J. {Schlegel}, {\'E}.~{Aubourg}, S.~{Bailey}, V.~{Bhardwaj},
  J.~R. {Brownstein} et~al., \emph{{Spectral Classification and Redshift
  Measurement for the SDSS-III Baryon Oscillation Spectroscopic Survey}},
  \href{https://doi.org/10.1088/0004-6256/144/5/144}{\emph{\aj} {\bfseries 144}
  (2012) 144} [\href{https://arxiv.org/abs/1207.7326}{{\ttfamily 1207.7326}}].

\bibitem{Hutchinson:2016AJ....152..205H}
T.~A. {Hutchinson}, A.~S. {Bolton}, K.~S. {Dawson}, C.~{Allende Prieto},
  S.~{Bailey}, J.~E. {Bautista} et~al., \emph{{Redshift Measurement and
  Spectral Classification for eBOSS Galaxies with the redmonster Software}},
  \href{https://doi.org/10.3847/0004-6256/152/6/205}{\emph{\aj} {\bfseries 152}
  (2016) 205} [\href{https://arxiv.org/abs/1607.02432}{{\ttfamily
  1607.02432}}].

\bibitem{Busca:2018arXiv180809955B}
N.~{Busca} and C.~{Balland}, \emph{{QuasarNET: Human-level spectral
  classification and redshifting with Deep Neural Networks}}, {\emph{arXiv
  e-prints} (2018) arXiv:1808.09955}
  [\href{https://arxiv.org/abs/1808.09955}{{\ttfamily 1808.09955}}].

\bibitem{Perez-Rafols:2020MNRAS.tmp.1953P}
I.~{P{\'e}rez-R{\`a}fols}, M.~M. {Pieri}, M.~{Blomqvist}, S.~{Morrison} and
  D.~{Som}, \emph{{Spectroscopic QUasar Extractor and redshift (z) Estimator
  SQUEzE I: Methodology}},
  \href{https://doi.org/10.1093/mnras/stz3467}{\emph{\mnras} (2020) }
  [\href{https://arxiv.org/abs/1903.00023}{{\ttfamily 1903.00023}}].

\bibitem{Perez-Rafols:2020MNRAS.tmp.1952P}
I.~{P{\'e}rez-R{\`a}fols} and M.~M. {Pieri}, \emph{{Spectroscopic QUasar
  Extractor and redshift (z) Estimator SQUEzE II: Universality of the
  results}}, \href{https://doi.org/10.1093/mnras/staa1786}{\emph{\mnras} (2020)
  } [\href{https://arxiv.org/abs/1911.04891}{{\ttfamily 1911.04891}}].

\bibitem{DESICollaboration:2016arXiv161100036D}
{DESI Collaboration}, A.~{Aghamousa}, J.~{Aguilar}, S.~{Ahlen}, S.~{Alam},
  L.~E. {Allen} et~al., \emph{{The DESI Experiment Part I: Science,Targeting,
  and Survey Design}}, {\emph{arXiv e-prints} (2016) arXiv:1611.00036}
  [\href{https://arxiv.org/abs/1611.00036}{{\ttfamily 1611.00036}}].

\bibitem{DESICollaboration:2016arXiv161100037D}
{DESI Collaboration}, A.~{Aghamousa}, J.~{Aguilar}, S.~{Ahlen}, S.~{Alam},
  L.~E. {Allen} et~al., \emph{{The DESI Experiment Part II: Instrument
  Design}}, {\emph{arXiv e-prints} (2016) arXiv:1611.00037}
  [\href{https://arxiv.org/abs/1611.00037}{{\ttfamily 1611.00037}}].

\bibitem{Ross:2012ApJS..199....3R}
N.~P. {Ross}, A.~D. {Myers}, E.~S. {Sheldon}, C.~{Y{\`e}che}, M.~A. {Strauss},
  J.~{Bovy} et~al., \emph{{The SDSS-III Baryon Oscillation Spectroscopic
  Survey: Quasar Target Selection for Data Release Nine}},
  \href{https://doi.org/10.1088/0067-0049/199/1/3}{\emph{\apjs} {\bfseries 199}
  (2012) 3} [\href{https://arxiv.org/abs/1105.0606}{{\ttfamily 1105.0606}}].

\bibitem{Guo:2019ApJ...879...72G}
Z.~{Guo} and P.~{Martini}, \emph{{Classification of Broad Absorption Line
  Quasars with a Convolutional Neural Network}},
  \href{https://doi.org/10.3847/1538-4357/ab2590}{\emph{\apj} {\bfseries 879}
  (2019) 72} [\href{https://arxiv.org/abs/1901.04506}{{\ttfamily 1901.04506}}].

\bibitem{DESI:2013internal}
DESI, \emph{Desi level 1 through level 3 requirements, science objectives,
  survey data set, instrument technical requirements},  nov, 2013.
\newblock DESI-doc-318.

\bibitem{Reid:2016MNRAS.455.1553R}
B.~{Reid}, S.~{Ho}, N.~{Padmanabhan}, W.~J. {Percival}, J.~{Tinker},
  R.~{Tojeiro} et~al., \emph{{SDSS-III Baryon Oscillation Spectroscopic Survey
  Data Release 12: galaxy target selection and large-scale structure
  catalogues}}, \href{https://doi.org/10.1093/mnras/stv2382}{\emph{\mnras}
  {\bfseries 455} (2016) 1553}
  [\href{https://arxiv.org/abs/1509.06529}{{\ttfamily 1509.06529}}].

\bibitem{Laurent:2017JCAP...07..017L}
P.~{Laurent}, S.~{Eftekharzadeh}, J.-M. {Le Goff}, A.~{Myers}, E.~{Burtin},
  M.~{White} et~al., \emph{{Clustering of quasars in SDSS-IV eBOSS: study of
  potential systematics and bias determination}},
  \href{https://doi.org/10.1088/1475-7516/2017/07/017}{\emph{\jcap} {\bfseries
  2017} (2017) 017} [\href{https://arxiv.org/abs/1705.04718}{{\ttfamily
  1705.04718}}].

\end{thebibliography}\endgroup

\end{document}